\documentclass[manuscript]{emulateapj}

\usepackage{color}
\usepackage{lipsum} 
\slugcomment{The Astrophysical Journal}

\shorttitle{X-raying the coronae of HD~155555}
\shortauthors{Lalitha et al.}

\begin{document}


\title{X-raying the coronae of HD~155555}


\author{S.~Lalitha$^1$, K.P.~Singh$^1$, S.A.~Drake$^2$,   $\&$ V. Kashyap$^3$}
\affil{1. Tata Institute of Fundamental Research, Homi Bhabha road, Mumbai 400005, India\\
2. NASA Goddard Space Flight Center, Code 662, Greenbelt, MD 20771, USA\\
3. Harvard-Smithsonian Center for Astrophysics, 60 Garden St., Cambridge, 02138, MA, USA}



\begin{abstract}

We present an analysis of the high-resolution \emph{Chandra}  observation of the multiple system, HD~155555 (an RS CVn type binary system, HD~155555~AB, and its spatially resolved low-mass companion HD~155555~C). This is an intriguing system which shows properties of both an active pre-main sequence star and a synchronised (main sequence) binary. We obtain the emission measure distribution, temperature structures, plasma densities, and abundances of this system and compare them with the coronal properties of other young/active stars. HD 155555 AB and HD 155555 C produce copious X-ray emission with log~$\mathrm{L}_\mathrm{x}$ of 30.54 and 29.30, respectively, in the 0.3-6.0 keV energy band. The light curves of individual stars show variability on timescales of few minutes to hours. We analyse the dispersed spectra and reconstruct the emission measure distribution using spectral line analysis. The resulting elemental abundances exhibit inverse first ionisation potential effect in both cases.  An analysis of He-like triplets yields a range of coronal electron densities $\sim$10$^{10}-10^{13}$ cm$^{-3}$. Since HD 155555 AB is classified both as an RS CVn and a PMS star, we compare our results with those of other slightly older active main-sequence stars and T Tauri stars, which indicates that the coronal properties of HD 155555 AB closely resemble that of an older RS CVn binary rather than a younger PMS star. Our results also suggests that the properties of HD 155555 C is very similar to those of other active M dwarfs.

\end{abstract}



\section{Introduction}
RS Canum Venaticorum (RS CVn) type stellar systems are close binaries containing late type evolved G/K stars \citep{hall_1976}. Typical orbital periods of these systems range from less than a day to a few months and, with a few exceptions such as $\lambda$ And, the rotation periods of the close components are spun up to match their orbital period due to the tidal synchronization forces. As a result of their rapid rotation and deep convective envelopes, RS CVns have powerful magnetic dynamos which generate enormous amounts of magnetic flux which, emerging into their outer layers, produces enhanced stellar activity. For example, RS CVns show optical variability due to the rotational modulation of an asymmetrical distribution of cool spots which can cover $>$~10\% of their surfaces. Studies of RS CVns in chromospheric lines such as Ca II H and K and H-alpha and in the ultraviolet and X-ray bands show that an enormous amount of energy (compared to stars like the Sun) is being released into their chromospheres and coronae, in both their typical (sometimes, called quiescent) and flaring states. Their intense X-ray activity was first detected by the HEAO-1 observatory \citep{walter_1981} and has been extensively studied since by subsequent missions such as \emph{Einstein, EXOSAT, ROSAT, ASCA, Chandra} and \emph{XMM-Newton}. The X-ray luminosities of RS CVn binaries are typically in the range of 30 $<$ log~$\mathrm{L}_\mathrm{x}$ $<$ 32 (ergs s$^{−1}$), corresponding to 10$^{-4}$ to 10$^{-3}$ of their bolometric luminosities, and log~$\mathrm{L}_\mathrm{x}$  values $>$~ 33  have been observed in large RS CVn flares (e.g. the Swift-detected flare of SZ Psc discussed by \citealt{drake_2015}).

Pre-main-sequence and early (zero-age) main-sequence solar- and sub-solar-mass stars have retained much of the angular momentum of their original protostellar predecessors, and thus have rotation rates from less than a day to $\sim$10 days, much in excess of the present solar rotation rate. As a result, just like active synchronized binaries with similar rotation rates, young stars have powerful chromospheres and coronae. As they evolve, their rotation rates, and consequently their activity levels, drop due to the loss of angular momentum in stellar winds. The coronae of very young (ages $<$~ 5 - 10 Myr) stars have been well-studied due to their locations in large nearby stellar associations and clusters, such as Tau-Aur and Orion OB1. Young, late-type pre-main sequence stars are luminous X-ray sources. This is indeed observed in T Tauri stars which are divided into two classes the Classic T Tauri stars (CTTS) and weak-line T Tauri stars (WTTS). The CTTS shows a strong H$\alpha$ emission lines and infra-red excess as a result of the accretion, whereas, WTTS display weak H$\alpha$ line and no infrared excess. \cite{guedel_2007} surveyed the Taurus molecular cloud and found evidence for lower X-ray luminosities in CTTS compared to WTTS. However, the X-ray emission of a fraction of CTTS, exhibits an additional soft, high-density component that has been attributed to a magnetically confined accretion column, e.g., the 8-Myr old TW Hya \citep{kastner_2002}. It has only been in the last 3 decades that the existence of a population of somewhat older (10 - 300 Myr) stars permeating the solar neighborhood in kinematically homogeneous stellar moving groups has been recognized, e.g., AB Dor \citep{rucinski_1982}, Speedy Mic \citep{bromage_1992}, etc. A few pre-main sequence stars also show evidence of soft extended X-ray emission close to the star that is thought to arise in accretion-related jet-like outflows, for example, DG Tau \citep{gudel_2005}, Z CMa \citep{stelzer_2009}, RY Tau \citep{skinner_2011}.

There has been some discussion in the literature as to whether the details of coronal and chromospheric activity in cool stars differs due to the underlying origin of their rapid rotation, i.e., spin-up in close binary systems vs. youth, but few definitive conclusions have been reached. A number of flares in active binaries observed in the radio have been observed by VLBI to have an extent similar to the separation between the two components (e.g., \citealt{little_1986}, suggesting that they may be related to inter-binary magnetic reconnection, but X-ray observations lack the milli-arcsecond spatial resolution needed for this type of analysis). There have been some indirect tests: e.g., \cite{singh_1996b}, compared 59 RS CVn binaries containing 2 late-type components to 29 Algol-type binaries most of which had only one late-type component, and found a tendency for RS CVns to be 3-4 times more X-ray luminous than Algols with the same orbital period, suggesting that there may be additional energy release in their coronae due to interconnecting magnetic field 
reconnections and/or disconnections. The most significant difference in the coronal properties of active RS CVns like AR Lac or active (pre-main sequence) binary system HD 98000 and active accreting (pre-main sequence) single stars like TW Hya are due to the presence of an accretion-related soft (T $\sim$ 2 - 4 MK) plasma component in CTTS; there can also be additional variability in the X-ray emission of CTTS due to varying absorption by circumstellar material and/or fluctuations in the accretion rate, e.g., \cite{flaccomio_2010}.

The intense coronal X-ray activity in the low mass RS CVn type binaries is a key characteristic in investigating the properties and evolution of low mass late-type stellar objects, spurring extensive studies with several X-ray satellites over the last few decades \citep{audard_2003}.  The Sun is usually considered a prototype of low mass cool stars, hence we often extrapolate the knowledge we have on the Sun to the other stars. The X-ray emission and the temperature of the coronal plasma of such stars are, however, at least a magnitude higher than the solar coronal emission indicating a fundamental difference in the dynamo powering the coronal activity. 
Most of the earlier studies, with a few exceptions, have been performed using low spectral resolution instruments, which were not able to provide very reliable information on coronal plasma. The unprecedented spectral resolution provided by X-ray gratings and large effective area of X-ray telescopes like \emph{Chandra} X-ray observatory and \emph{XMM-Newton}, allows us to resolve individual lines thus providing plasma diagnostics of coronae of stars other than the Sun.  In particular, an updated picture of coronal abundance anomalies based on  observations of HR 1099 confirmed its depletion in the abundance of  low first ionisation potential (FIP) elements in comparison to the high-FIP elements  \citep{brinkman_2001, drake_2001, guedel_2001}. Later, \cite{audard_2002} showed that stars with different activity level show a continuum of behaviour from the FIP effect to the inverse FIP (iFIP) effect. 

Spectral diagnostics based on a large number of prominent emission lines can be used to derive the temperatures, density structure  of emitting plasma, emission processes and ionisation properties of the coronal X-ray plasma. Furthermore, based on the electron density and coronal temperatures we can infer the pressure and the volume of the emitting plasma. Since stellar coronal structures are spatially unresolved, unlike the solar coronal loops, by combining all the above results one can also derive useful information about the coronal loop properties \citep{reale_1997, guedel_2009, lalitha_2011}. 

Finally, we can compare the high-resolution X-ray spectroscopic data with that of other low mass stars with different ages which are crucial for understanding how the stellar activity and coronal properties change during these different evolutionary phases \citep{ball_2005, robrade_2005}.

Here, we report the first high spectral resolution observations of the HD~155555 system with \emph{Chandra} HETG with the objective of getting a better insight into the hottest part of the outer stellar atmosphere of the active stars in HD~155555 system. Although stellar activity has been observed and studied for decades, the exact mechanism controlling the magnetic activity in low-mass stars are still not understood. The fundamental parameters must be derived for a range of stars in order to provide the basis for a theoretical understanding. Our goal is to use the X-ray light curves and spectra of an active RS CVn system to address several issues: monitor the flares, determine the emission measure and temperature distribution, determine abundances, probe the density diagnostics and compare to other stars in different evolutionary states. This paper is organised as follows: in \S~\ref{sec:sys} we discuss the properties of HD~155555~system; \S~\ref{sec:obs} presents the details of the X-ray observation and the methods adopted for the data analysis; in \S~\ref{sec:anal_ab} we describe the procedures used to obtain the 
emission measures, the abundances and the electron densities of HD~155555~AB. In \S~\ref{sec:anal_c}, we analyse the time variability and spectral properties of HD~155555~C. We discuss the implications of the analyses of both HD~155555~AB and the companion HD~155555~C in \S~\ref{sec:disc} and summarise our conclusions in \S~\ref{sec:sum}.

\section{The HD 155555 system}\label{sec:sys}
HD~155555 (V824 Ara), is a young ($\approx$18~Myr, \citealt{strassmeier_2000}) RS CVn binary which also shows some characteristics common to the pre-main sequence stars. It is a triple star system consisting of cool stars: G5IV, a K0IV and an M3V star. Two of the three stars (G and K stars) are in a close binary system with an orbital period of 1.68d \citep{strassmeier_2000} and will together be  referred to as HD~155555~AB,~henceforth. The M dwarf companion LDS~536~B is 33$\arcsec$ away from the primary close-in binary system, and henceforth will be referred to as HD~155555~C. \cite{pasquini_1991} suggested that HD~155555~AB and HD~155555~C are part of a young disk which is in agreement with the high lithium abundance in both the components. Later, based on the Hipparcos satellite the distance to the HD 155555 system has been determined to be 31 pc \citep{strassmeier_2000}. The properties of the stars in HD~155555 system are summarised in Table~\ref{tab:physicalparam}.

HD~155555~AB exhibits all the characteristics nominally attributed to the RS~CVn class of stars, viz., strong \ion{Ca}{2} H and K emission due to  chromospheric activity and photometric variability as a result of spot migration \citep{cutispoto_1990}. \cite{dunstone_2008_2} have shown that both the components of the HD~155555 system are particularly interesting because it is an unusual RS~CVn in which the close-in binaries have become synchronised rotators as they evolved towards the main sequence; most RS~CVn systems synchronise much later in their lifetimes as one/both stars evolve away from the main sequence. Thus, without the binarity factor, HD~155555 system would still be active, since they are rapidly rotating cool stars.  The binarity of this system has, however, set the orbital period to 1.68d which is shorter than if this were a single star. It is likely that while both age and binarity have worked together to produce activity, in this case, the latter has determined the exact period.

\begin{table}[!ht]
\centering
\caption{\label{tab:physicalparam} Physical parameters of the HD~155555 system.}
\begin{tabular}{lcllccccc}
\tableline
Parameter& HD~155555A & HD~155555B& HD~155555~C\\
\tableline
\vspace{0.1em}
V$_{mag}$&6.72 & &12.82\\
T [K] & 5500$\pm$100&5050$\pm$150 &3300$\pm$150\\
Spectral type &  G5IV & K0IV & M3V\\
P [days] & 1.68 & 1.68 &\\
Inclination i &52.5$^0\pm$2.5&52.5$^0\pm$2.5&50$^0\pm5$\\
V$sin$i [km~s$^{-1}$] &36.8$\pm$1.0& 33.7$\pm$1.5&$\leq$3\\
R$sin$i [R$_{\odot}$]&1.55$\pm$0.18&1.21$\pm$0.25&\\
M$sin$i [M$_{\odot}$]&1.10$\pm$0.01&1.00$\pm$0.01&\\
log~$\mathrm{L}_\mathrm{bol}$ [erg~s$^{-1}$]&33.89&&32.26\\  
\tableline

\end{tabular}

\footnotesize{References. (a) \cite{pasquini_1989}; (b) \cite{strassmeier_2000}.}\\

\end{table}


The stars in HD~155555~AB probably exhibit substantial differential rotation which is a key factor in the generation of stellar magnetic fields. A very high level of activity has also been reported from the visual companion HD~155555~C \citep{martin_1995}. The X-ray emission from HD~155555~AB was first reported by \textit{HEAO-1}~A2 low energy experiment \citep{walter_1980}. It was found to be a relatively strong X-ray source with log~$\mathrm{L}_\mathrm{x}$ $\approx$1.0$\times$10$^{31}$ erg~s$^{-1}$, yielding a log~$\frac{\mathrm{L}_\mathrm{x}}{\mathrm{L}_\mathrm{bol}}$ $\sim$-2.9 in 0.5-3~keV energy range, making it a very active binary system. It was subsequently observed by several other X-ray satellites and a summary of observed X-ray fluxes and the corresponding luminosities is given in Table~\ref{tab:xflux}. The low X-ray flux observed by \textit{EXOSAT} LE could be due to poor calibration. HD~155555~C was discovered serendipitously during EXOSAT observations of HD~155555~AB \citep{pasquini_1989}. The X-ray luminosity of HD~155555~C was found to be $\sim$3$\times$10$^{29}$ erg~s$^{-1}$ in 0.04-2.0 keV energy band, yielding a log~$\frac{\mathrm{L}_\mathrm{x}}{\mathrm{L}_\mathrm{bol}}$$\sim$-2.8, suggesting that HD~155555~C is also a very active young star.

\begin{table}[!ht]
\centering
\caption{\label{tab:xflux}  The X-ray luminosity of the HD~155555~AB system.}
\begin{tabular}{lcccccccc}
\tableline
Instruments  & Unabsorbed F$_X$ & $\mathrm{L}_\mathrm{x}$&Ref.\\
            & $10^{-11}$ erg s$^{-1}$ cm$^{-2}$ & 10$^{30}$ erg s$^{-1}$& \\
\tableline

\emph{HEAO-1 A2} &9.0& 10.31 &(a)\\
(0.5-3.0 keV)&&&\\
\emph{EXOSAT} & 0.9-1.1& 1.03-1.26 &(b)\\
(0.04-2.0 keV)&&&\\
\emph{ROSAT PSPC} & 3.8& 4.35&(c)\\
(0.1-2.4 keV)&&&\\
\emph{ROSAT HRI} & 3.7-4.6& 4.23-5.26&(c)\\
(0.1-2.4 keV)&&&\\
\emph{XMM-Newton} & 5.7&6.52&(d)\\
(0.2-12.0 keV)&&&\\
\emph{Chandra} HETG &&&\\ 
(0.3-6.0 keV)& 2.8-3.6&3.20-4.12&(e) \\
(0.5-5.0 keV)&2.4-3.0&2.74-3.43&(e)\\
&&& \\

\tableline

\tableline

\end{tabular}

\footnotesize{References. (a) \cite{walter_1980}; (b) \cite{barstow_1987}; (c) Using 
XSPEC v13.0 and WebPIMMS v4.7 and assuming a coronal temperature of 10~MK, 
we estimated ECF$_{PSPC}$=7.9$\times$10$^{-12}$ erg s$^{-1}$ cm$^{-2}$ and 
ECF$_{HRI}$=2.5$\times$10$^{-11}$ erg s$^{-1}$ cm$^{-2}$}; (d) observed during XMM-Newton slew survey 
and converted to flux using the flux limit 1.3$\times$10$^{-11}$ erg s$^{-1}$ cm$^{-2}$; (e) present work.\\

\end{table}


\section{The \emph{Chandra} data}\label{sec:obs}
The HD~155555 system was observed for $\sim$94~ks starting on 2002 July 10 11:27 UT, using the HETG \citep{weisskopf_2002} and Advanced Camera for Imaging Spectroscopy \citep[ACIS,][]{garmire_2003} on-board \emph{Chandra} X-ray observatory (ObsID 2538). The data obtained from the public \emph{Chandra} Data Archives were reprocessed following the \emph{Chandra} Interactive Analysis of Observations (CIAO ver 4.5)\footnote{see http://cxc.harvard.edu/ciao/threads/} science threads, and XSPEC V13.0 \citep{arnaud_1996} was used for spectral analysis.

\begin{figure}[!ht]

\centering
\includegraphics[width=8cm, height=5cm]{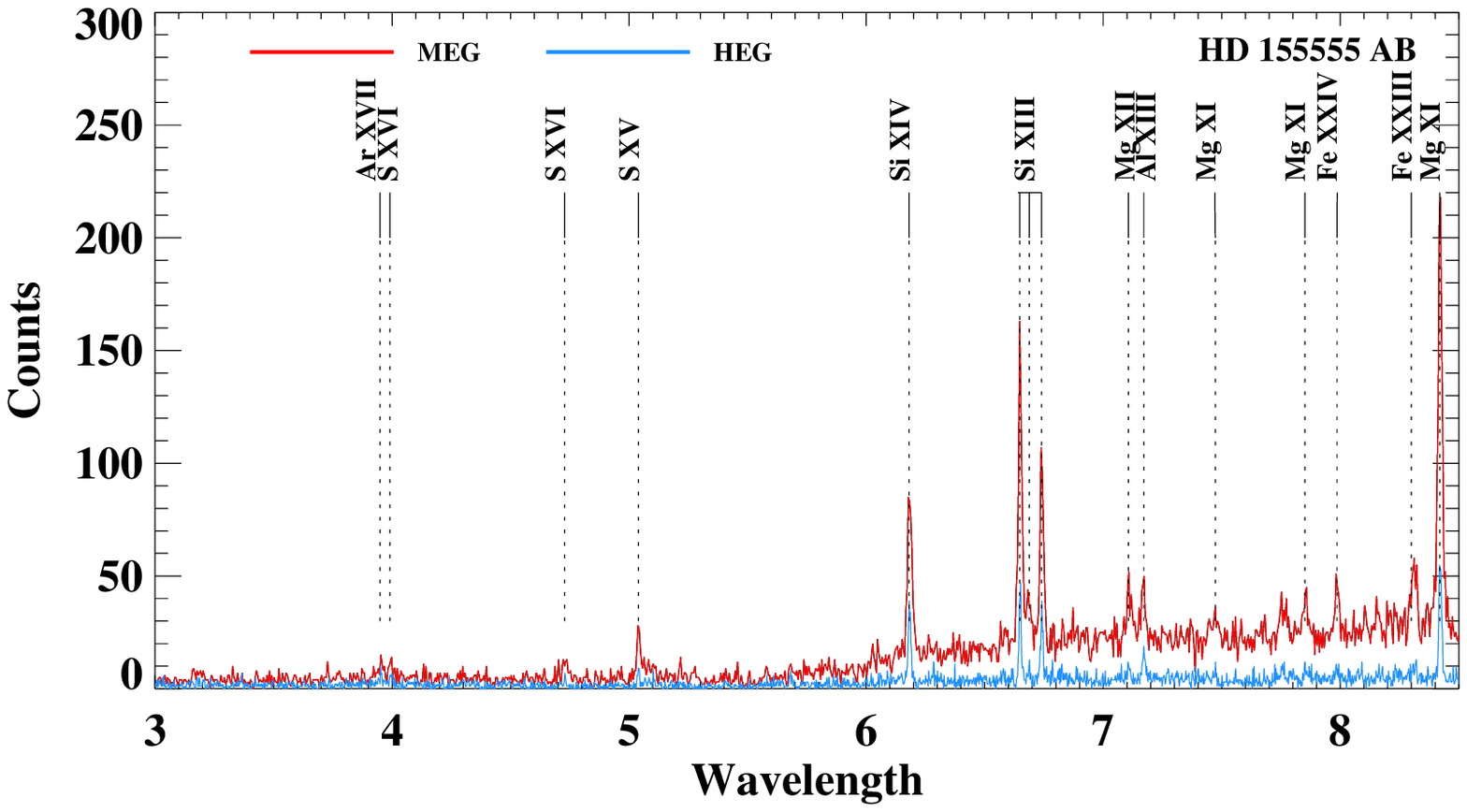}
\includegraphics[width=8cm, height=5cm]{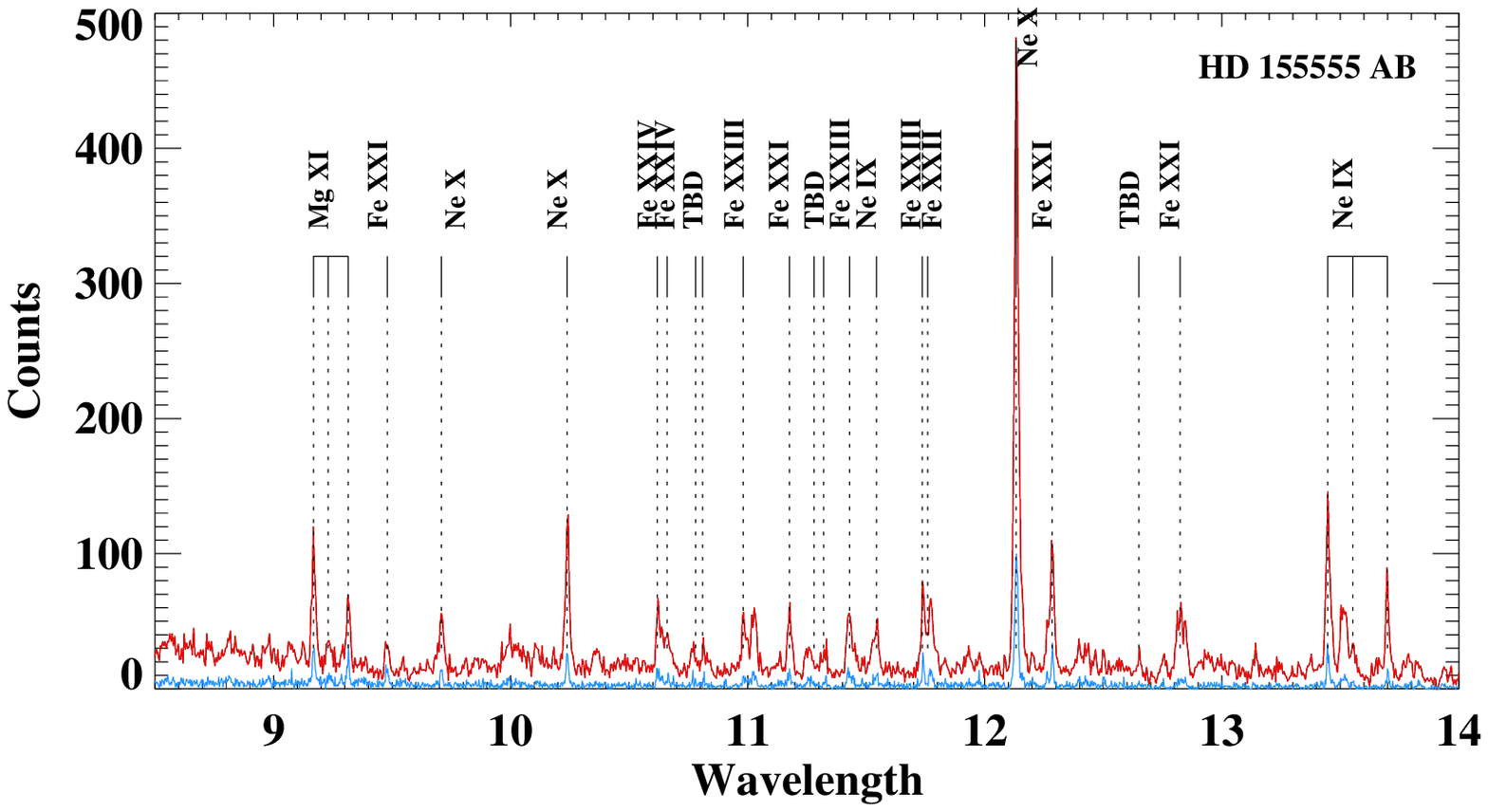}
\includegraphics[width=8cm, height=5cm]{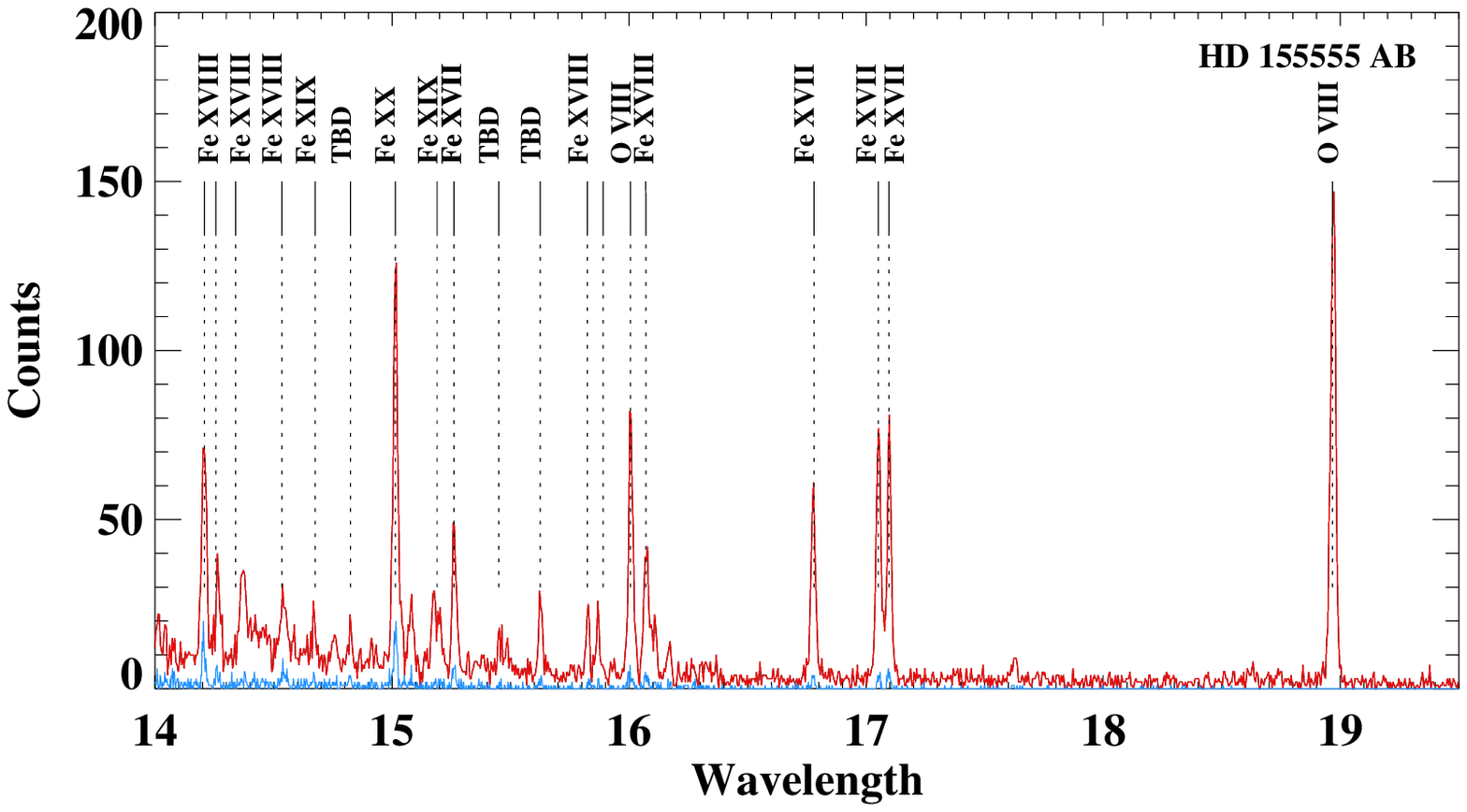}
\includegraphics[width=8cm, height=5cm]{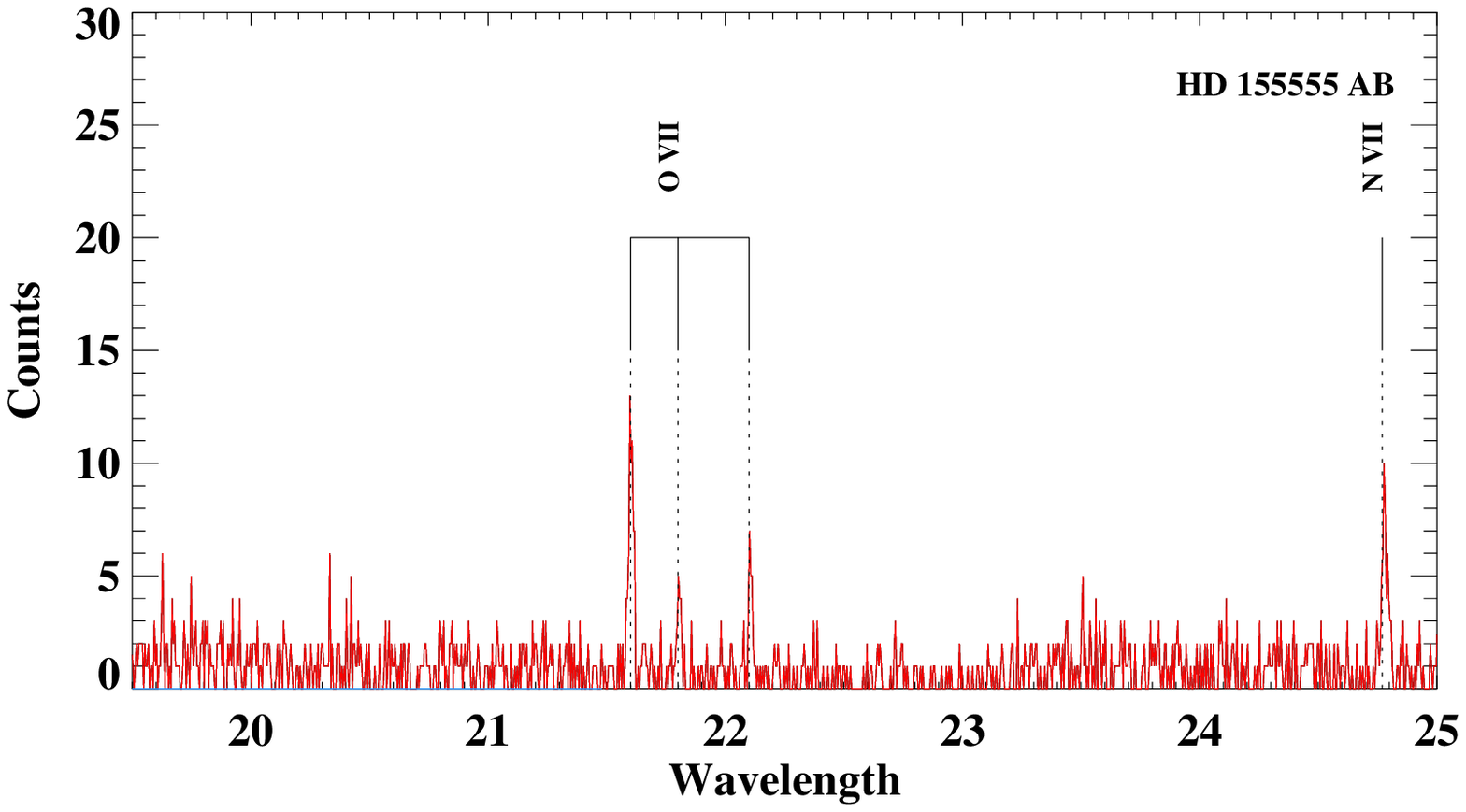}
\caption{\label{fig:abspec} Combined +/- order spectrum of 
HD~155555~AB for the entire observation binned to 0.005$\textrm{\AA}$ for MEG (red) and 
0.0025$\textrm{\AA}$ for HEG (blue). The prominent lines are labeled.}

\end{figure}

\begin{figure*}[!ht]
\centering
\includegraphics[width=8cm, height=5cm]{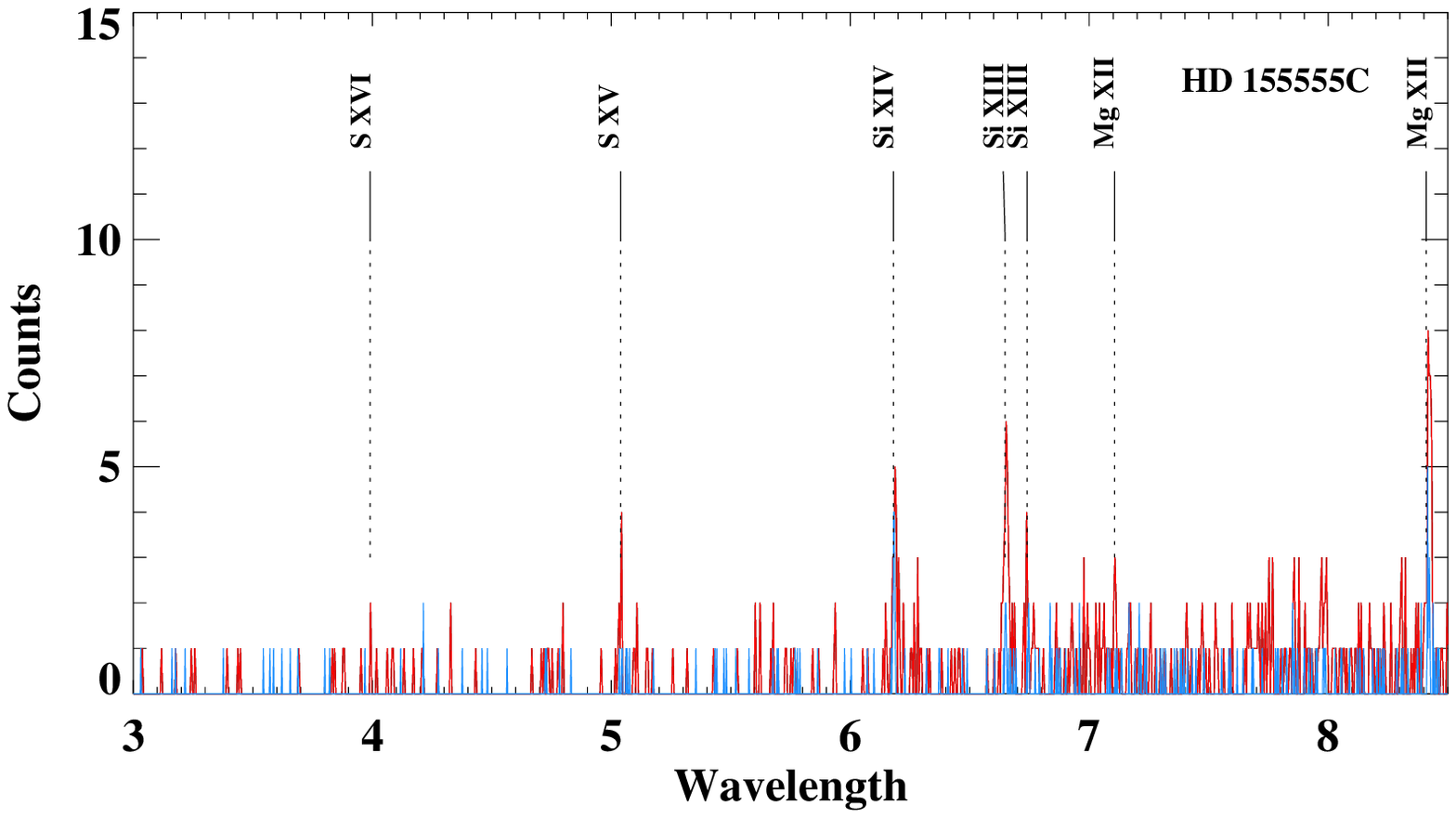}
\includegraphics[width=8cm, height=5cm]{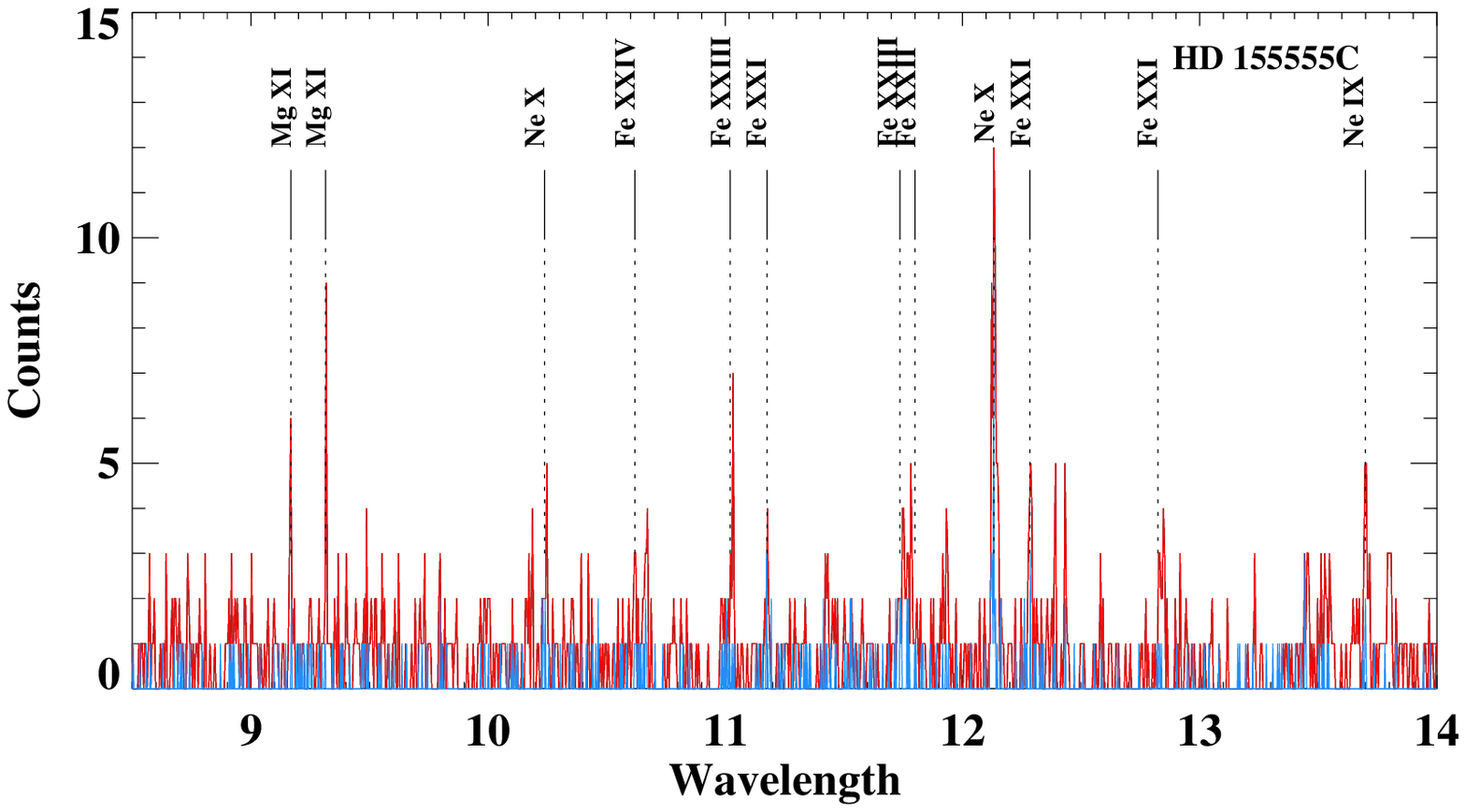}
\includegraphics[width=8cm, height=5cm]{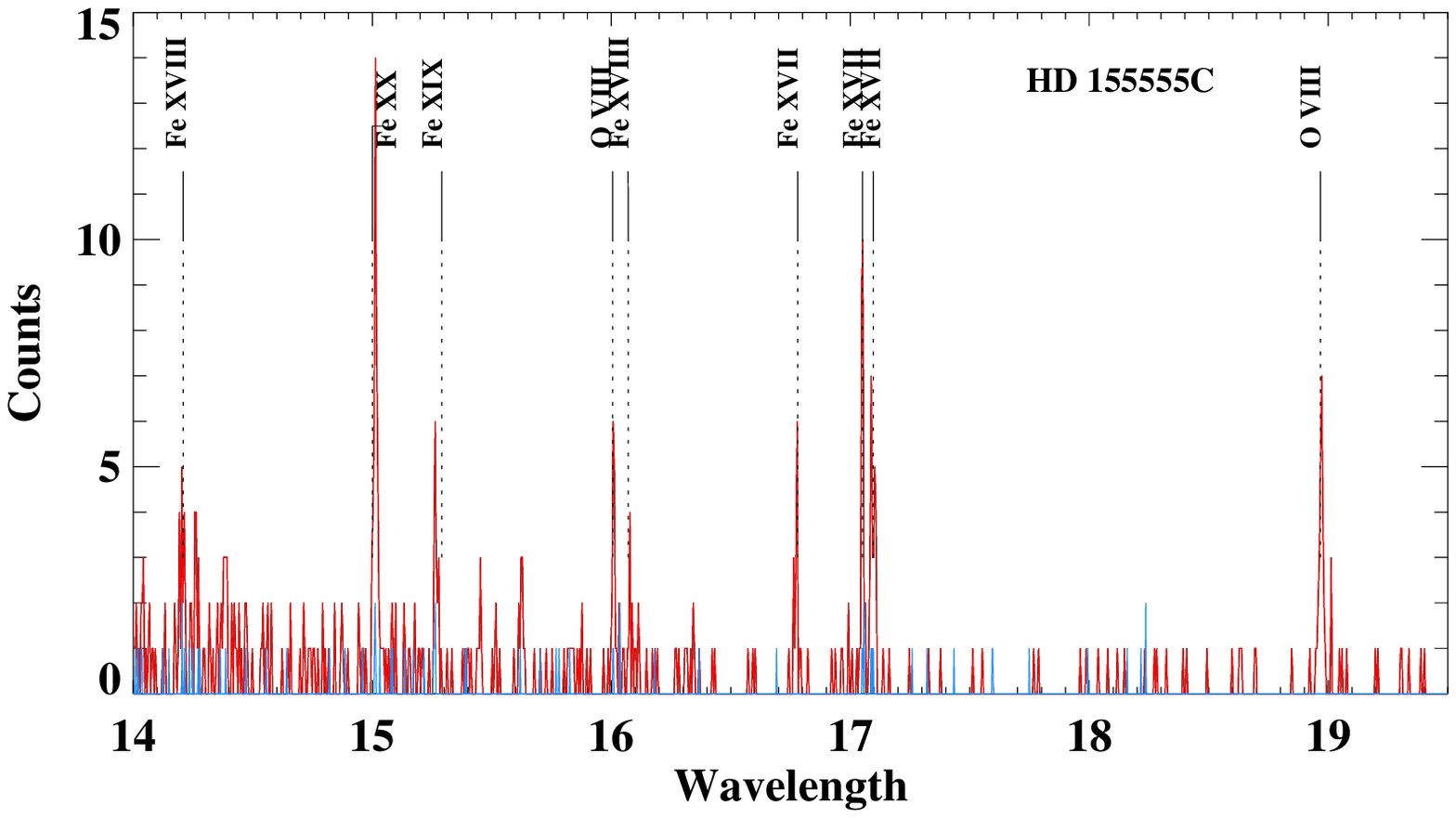}
\includegraphics[width=8cm, height=5cm]{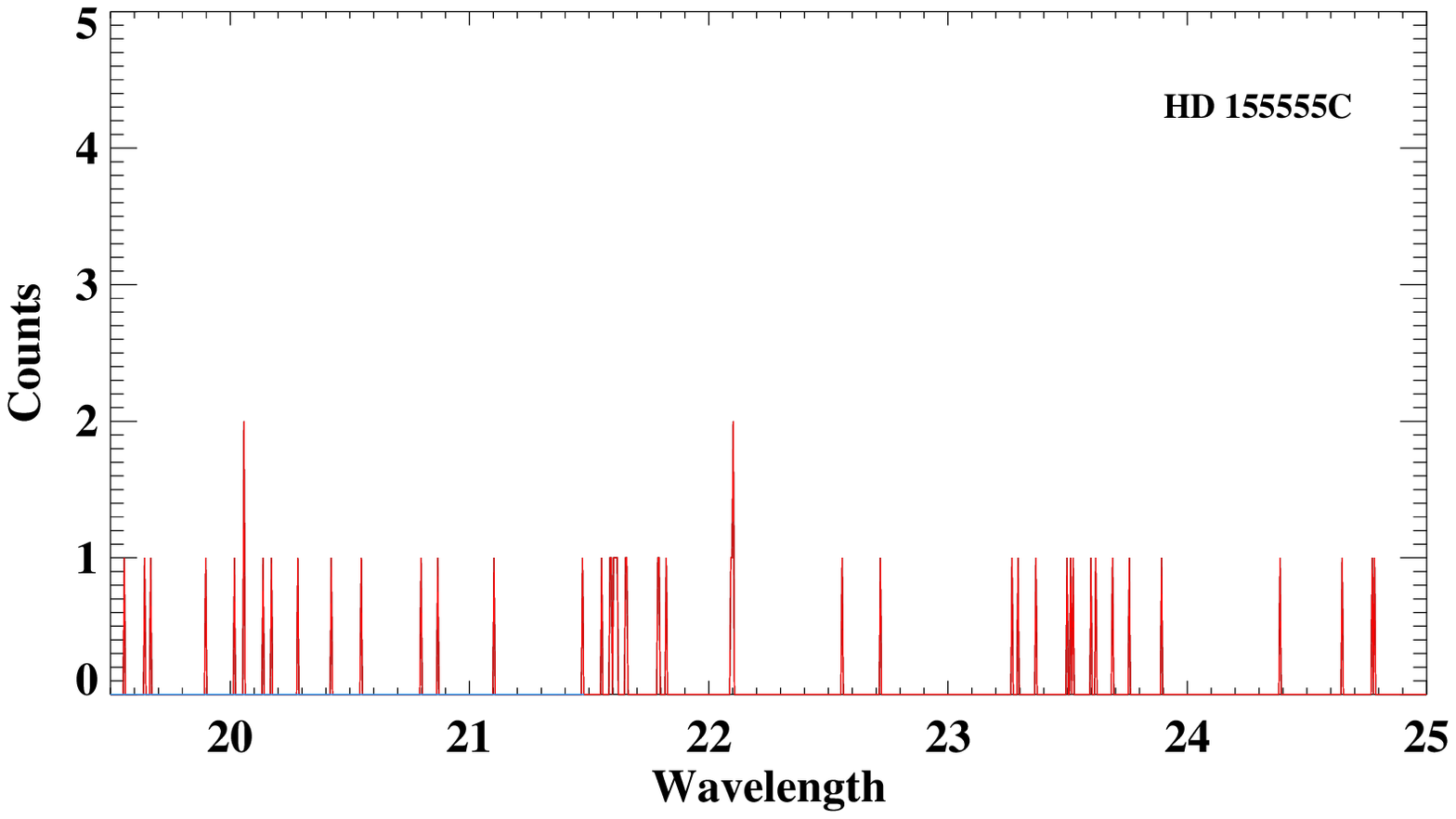}
\caption{\label{fig:cspec} Combined +/- order spectrum of 
HD~155555~C for the entire observation binned to 0.005$\textrm{\AA}$ for MEG (red) and 
0.0025$\textrm{\AA}$ for HEG (blue). The prominent lines are labeled.}

\end{figure*}

To optimize the signal-to-noise ratio in the spectrum, CIAO includes position-dependent PHA filtering of the data. We used the TGEXTRACT tool available in the CIAO to extract the -1 and +1 grating orders of HEG and MEG, and used the \textit{add$\_$grating$\_$orders} script to sum the orders to improve the counts statistics. We used the default binning of 0.005$\textrm{\AA}$ for MEG and 0.0025$\textrm{\AA}$ for HEG. The background spectrum was extracted from source-free symmetrical regions near the source region, with each region having an area of the order of few times that of the source area. We computed the net spectrum of HD~155555~AB and HD~155555~C by subtracting the background spectrum. The spectra were then scaled to the area of the source region, to get the source spectrum for both +1 and -1 orders. The positive and negative orders spectra for each of the sources were then combined to get the final spectrum. The co-added negative and positive first order MEG and HEG spectrum of HD~155555~AB and HD~155555~C for the entire observation are shown in Figure~\ref{fig:abspec} and ~\ref{fig:cspec}. There are numerous well resolved lines with high signal-to-noise ratio present in the spectrum and these have been labelled in the figures. 


Furthermore, we re-ran the CIAO data reduction pipeline and extracted the zeroth-order spectrum of both HD~155555~AB and HD~155555~C. The zeroth-order spectrum is a low-resolution spectrum created by the photons that pass through the grating without being diffracted. The net count rate for HD~155555~AB and HD~155555~C zeroth order spectrum is 0.23$\pm$1.58$\times$10$^{-3}$ and 3.96$\times$10$^{-2}$$\pm$6.54$\times$10$^{-4}$ counts s$^{-1}$, respectively. Based on the observation of HD~155555 system, PIMMS predicts a large pile-up fraction of $>$20$\%$; 
hence, we chose to use only the dispersed spectrum for our analysis.

\section{Analysis and results: HD~155555~AB}\label{sec:anal_ab}

\subsection{Variability}
\begin{figure}[!ht]
\centering
\includegraphics[width=8.5cm, height=6cm]{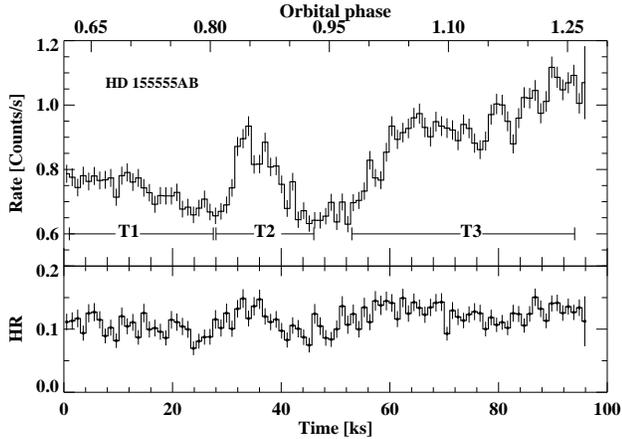}
\caption{\label{fig:lc} Top panel: Background subtracted X-ray light curve in 0.4-6.0 keV energy band and the corresponding orbital phase assuming an orbital period of $\sim$1.68 days for HD~155555~AB. Bottom panel: The time variability of the ratio of number of counts in the 2.0-6.0 keV band to the number of counts in the 0.4-2.0 keV band for HD~155555~AB.} 

\end{figure}

We produced light curves from the event list using the dispersed events for the full $\sim$94~ks observing period. In Figure~\ref{fig:lc} (top panel), we show the X-ray light curve in the 0.4-6~keV energy band. In this figure, we also show the orbital phase corresponding to our observation using the ephemeris by \cite{strassmeier_2000} $HJD = 2,446,997.910 + 1.6816463 (\pm0.0000003) \times E$. The light curve shows a significant variability and  has been divided into three distinct time bins marked as T1, T2 and T3, thus isolating the various features noticed in the light curve. The time bin T1 spans from 0-25~ks which are free of any large variation; the time bin T2 spans between 25-45~ks and shows a typical flare-like feature with a long and sharp rise in the flux and a slow exponential decay; the time bin T3 shows a broad event starting at $\sim$45 ks and lasting until the end of the observations.   

The smooth rise in X-ray emission (in time bin T3) between T=50 and 60~ks (corresponding to rotation phase $\phi\sim$1.0) is followed by a plateau which could be due to the rotation of one of the two stars in HD~155555~AB crossing the visible hemisphere of a coronal active region. If it lasts for $\sim$10~ks, and assuming that the active region occurs near the equator and neglecting the inclination and differential rotation, we estimate a size of $\sim$25$^{\circ}$ ($\frac{10 ~ks}{P_{rot} ~ks}\times 360^{\circ}$ $\approx$25$^{\circ}$) for the active  region. However, Doppler imaging of HD 155555 AB carried out by \citep{strassmeier_2000} reveals inhomogeneities at higher latitudes this would force the active region dimension to be bigger than $\sim$25$^{\circ}$, hence this value is essentially a lower limit. Similarly, there may be a rise between 80 and 95~ks, but it is not as clear as the previous rise and  it may just be due to small-scale variation occurring at $\sim80$, $\sim85$ and $\sim$92~ks.

\begin{figure}[!ht]
\centering
\includegraphics[width=8.5cm, height=6cm]{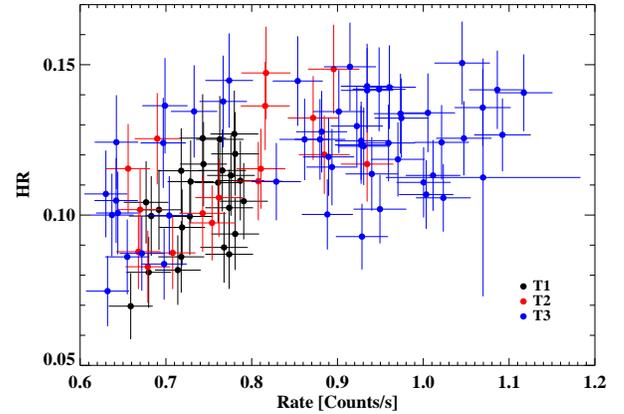}
\caption{\label{fig:hrlc} Hardness ratio of HD~1555555~AB as function of the background subtracted X-ray count rate in 0.4-6.0 keV energy band. The black, red and blue filled circles refer to the T1, T2 and T3, respectively, (see text for details).} 

\end{figure}

We searched for spectral variations caused by flare heating events (e.g., events seen during the phase T2) by defining a hardness ratio (HR) as the ratio of the number of counts in the hard band (2.0-6.0~keV) to the number of counts in the soft band (0.4-2.0 keV). The hardness ratio variations of HD~155555~AB are shown in the bottom panel of Figure~\ref{fig:lc}. In Figure~\ref{fig:hrlc}, we plot the HR vs the count rate of the T1 (black filled circles), T2 (red filled circles) and T3 (blue filled circles). 
We carried out a non-parametric correlation test on the HR and count rates for each of the time bins. The rank correlation co-efficient with two-sided significance values are given in Table~\ref{tab:cor}. The small p-value indicates a correlation with 2.5-3$\sigma$ probability between the HR and the count rate for the time bin T2 with a flare. The correlation observed in the other time bins is of lower significance. We have carried out parametric regression analysis and obtained the slopes for each of the time bins and shown in (column~4, Table~\ref{tab:cor}). 

\begin{table}[!ht]
\centering
\caption{\label{tab:cor} Results of correlation test and regression analysis performed on the HR and the count rate for each of the time bins.}
\begin{tabular}{lcccccccc}
\tableline
Time bin& \multicolumn{2}{c}{Spearman's correlation}& \multicolumn{1}{c}{Linear regression}\\
            & $\rho$ & p-value& slope\\
\tableline            
T1&0.44&0.025&0.19$\pm$0.07\\
T2&0.66&0.004& 0.14$\pm$0.04\\
T3&0.35&0.010&0.05$\pm$0.02\\

\tableline

\tableline

\end{tabular}

\end{table}

%

\begin{figure}[!ht]
\centering
\includegraphics[width=8.5cm, height=6cm]{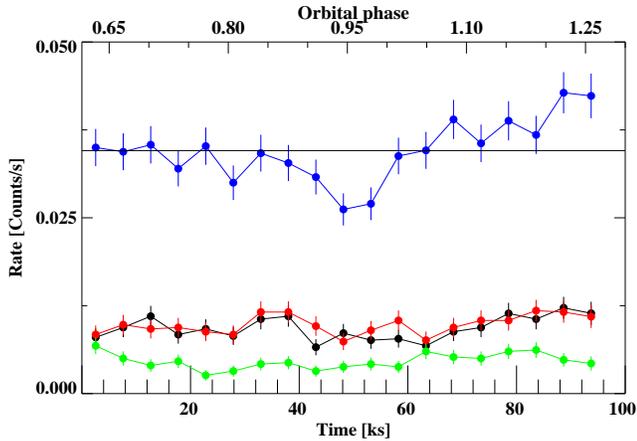}
\caption{\label{fig:line} X-ray light curve from combined MEG +1 and -1 order counts for HD~155555~AB in the \ion{Mg}{11} at 8.41\AA ~(green), \ion{Ne}{10} at 12.13\AA~(blue), \ion{Fe}{17} at 15.01\AA ~(red) and \ion{O}{8} at 18.97\AA ~(black). The mean count rate for \ion{Ne}{10} line is indicated by a thick vertical black line.}

\end{figure}

We have also examined the prominent line features in the spectra for variability. We find that the \ion{Ne}{10} line feature at 12.13 $\textrm{\AA}$ shows evidence of variability. Its light curve is shown in Figure~\ref{fig:line}. We tested other strong lines like \ion{Mg}{11} at 8.41$\textrm{\AA}$, \ion{Fe}{17} at 15.01 $\textrm{\AA}$, and \ion{O}{8} at 18.97 $\textrm{\AA}$ but they show very marginal variation or no significant variation. The detection of a dip in \ion{Ne}{10} seen close to $\phi\sim$1.0 and coinciding with the dip in the total intensity in Fig.~\ref{fig:lc} could indicate that moderately cool gas gives rise to this feature. The rise in the \ion{Ne}{10}  flux towards the end of the observation probably indicates the presence of a large active region on the surface of the star moving into the line of sight.

%

\subsection{Spectroscopic behaviour}\label{sec:spec}

We  divided the source 
spectra into three time-intervals as mentioned above and carried out a detailed spectral analysis for each of these intervals to investigate the temperature, the emission measure, the abundance and the density variation of the coronal plasma in HD~155555~AB. The analysis was carried out by global fitting using XSPEC as well as line by 
analysis of prominent lines using PINTofALE.

\subsubsection{XSPEC fitting} \label{sec:xspecab} \hfill \break
We have characterised the thermal structure and the elemental abundance composition of the coronae of HD~155555~AB using the simultaneous global fitting to the HEG and MEG spectra. The temperatures and abundances relative to the solar values \citep{grevesse_1998} were determined by simultaneous iterative XSPEC fit to the HEG and MEG spectra with Variable Astrophysical Emission Code (VAPEC) for thermal optically thin plasma models \citep{smith_2001}. We determined the number of VAPEC components with different temperatures by adding a new component until the addition of the next component did not improve the fit. Assuming that the corona of a star is not isothermal and has multi-temperature components, we used two-temperature VAPEC models. This produced a $\chi_{min}^2\sim$ 611 (DOF$\sim$711), $\chi_{min}^2\sim$ 412 (DOF$\sim$591), and $\chi_{min}^2\sim$ 1419 (DOF$\sim$1320) for time bins T1, T2 and T3, respectively, showing that 2T VAPEC models are a good fit to the data. We then added a new temperature component 
which reduced the minimum $\chi^2$ further to  $\chi_{min}^2\sim$ 520 (DOF$\sim$709), $\chi_{min}^2\sim$ 368 (DOF$\sim$589), and $\chi_{min}^2\sim$ 1245 (DOF$\sim$1318) for time bins T1, T2 and T3, respectively. The F-test produces a p-value of 0.017, 0.0913 and 0.0096 for time bins T1, T2 and T3, respectively. These p-values could under-state the evidence for the extra component by factors of 2 \citep{protassov_2002}, so we use a 3T model. Adding a 4th temperature component yields an F-test p-values 0.48, 0.47 and 0.50 for time bins T1, T2 and T3, respectively, which are unambiguously large. Hence, we use the 3T VAPEC models to describe of the dispersed spectra of HD~155555~AB. 
We allowed the abundance of all the elements to vary freely for each of the time bins. However, we fixed them to the same abundance values for each of the temperature components in the 3T VAPEC. For example, in the case of time bin T1 we allowed the O, Ne, Mg, Al, Si, S, Ar, Fe and Ni abundances vary freely while tying all the temperature components to the same abundance value for a given element. We, thus obtain a unique solution for the abundances for each time bin separately and also the temperatures, and emission measures.The best-fit temperatures and the emission measures for time intervals T1, T2 and T3 are presented in Table~\ref{tab:specparam}, whereas 
the best-fit abundance values are listed in Table~\ref{tab:abundab}. The unabsorbed X-ray flux of HD~155555~AB in 0.3-6.0 keV energy band is in the range $\sim2.8-3.6 \times 10^{-11}$ erg s$^{-1}$ cm$^{-2}$, implying L$_X$=3.20-4.12$\times$10$^{30}$ erg s$^{-1}$. 


\begin{table}[!ht]
\centering
\caption{\label{tab:specparam}  Best-fit spectral parameters obtained for time intervals T1, T2, T3 from iterative fitting of dispersed X-ray spectra for HD~155555~AB data using 3T VAPEC plasma models. See Table~\ref{tab:abundab} for the best-fit elemental abundances.}
\begin{tabular}{lccccccc}
\tableline
Parameters &  T1  &T2 &T3 \\
\tableline
log t$_1$ [K] & 6.72$\pm$0.03& 6.74$\pm$0.07&6.64$\pm$0.05\\
EM$_1$ [cm$^{-3}$] & 52.92$^{+0.10}_{-0.11}$& 52.94$^{+0.12}_{-0.24}$&52.93$^{+0.02}_{-0.19}$\\
log t$_2$ [K] & 7.03$\pm$0.05& 7.01$\pm$0.07&6.94$\pm$0.02\\
EM$_2$ [cm$^{-3}$] & 53.10$^{+0.13}_{-0.11}$ & 53.06$^{+0.14}_{-0.11}$&53.10$^{+0.06}_{-0.05}$\\
log t$_3$ [K] & 7.34$\pm$0.10 & 7.26$\pm$0.07&7.23$\pm$0.01\\
EM$_3$ [cm$^{-3}$] & 52.93$^{+0.20}_{-0.10}$ & 53.04$^{0.10}_{-0.22}$&52.28$^{+0.23}_{-0.44}$\\
red. $\chi^2$ & 0.73 &0.62& 0.94\\
DOF &709&589&1316\\
log~$\mathrm{L}_\mathrm{x}$ [erg\,s$^{-1}$]&30.51&30.52&30.61\\
log~$\frac{\mathrm{L}_\mathrm{x}}{\mathrm{L}_\mathrm{bol}}$&-3.37&-3.36&-3.27\\

\tableline

\end{tabular}

\footnotesize{Note: the errors are estimated with 90\% confidence limit.}\\

\end{table}


\subsubsection{Line based analysis}\label{sec:lineab}
We identified 
the individual line features on the basis of CHIANTI version 4.02 (Dere et al. 2001) database. The identified lines and their corresponding fluxes are listed in Appendix. 
In Figure~\ref{fig:abspec}, both the MEG and HEG spectra are shown. The lines in the MEG spectra are much brighter than in the HEG spectra. Therefore, we 
consider only the lines seen in the MEG for further analysis. 

We derive the differential emission measure (DEM) from the measured line fluxes using Markov Chain Monte Carlo (MCMC) analysis method of \cite{kashyap_1998} as implemented in the PINTofALE\footnote{see http://hea-www.harvard.edu/PINTofALE/} software package. The DEM gives the amount of plasma as a function of temperature. The MCMC analysis gives an estimate of emission measure distribution over a pre-selected range of temperatures with a DEM estimate for each of the temperature bins. We used a temperature grid with $\delta$log T=0.05 with a range of 6.2$<$log T[K]$<$7.4. All strong emission lines which are reasonably isolated (minimum separation $\sim0.01\AA$) were used for the DEM reconstruction. We derived the abundance relative to the iron for all the elements for which at least one line has been identified and included in the selection. The elemental abundances were obtained along with reconstructed DEM distribution. For this, we used only the iron lines from our selection to obtain the initial DEM distribution and abundance of Fe. We then added the lines of other elements one by one gradually extending to the temperature range where the DEM is constrained. At each iteration, we obtained a new DEM distribution and the abundance relative the Fe abundance. Once the DEM distribution has been calculated the Fe abundance is determined by comparing the observed continuum with the predicted continuum based on different metallicity values (see \citealt{argiroffi_2003}). 

\begin{figure}[!ht]
\centering
\includegraphics[width=0.45 \textwidth]{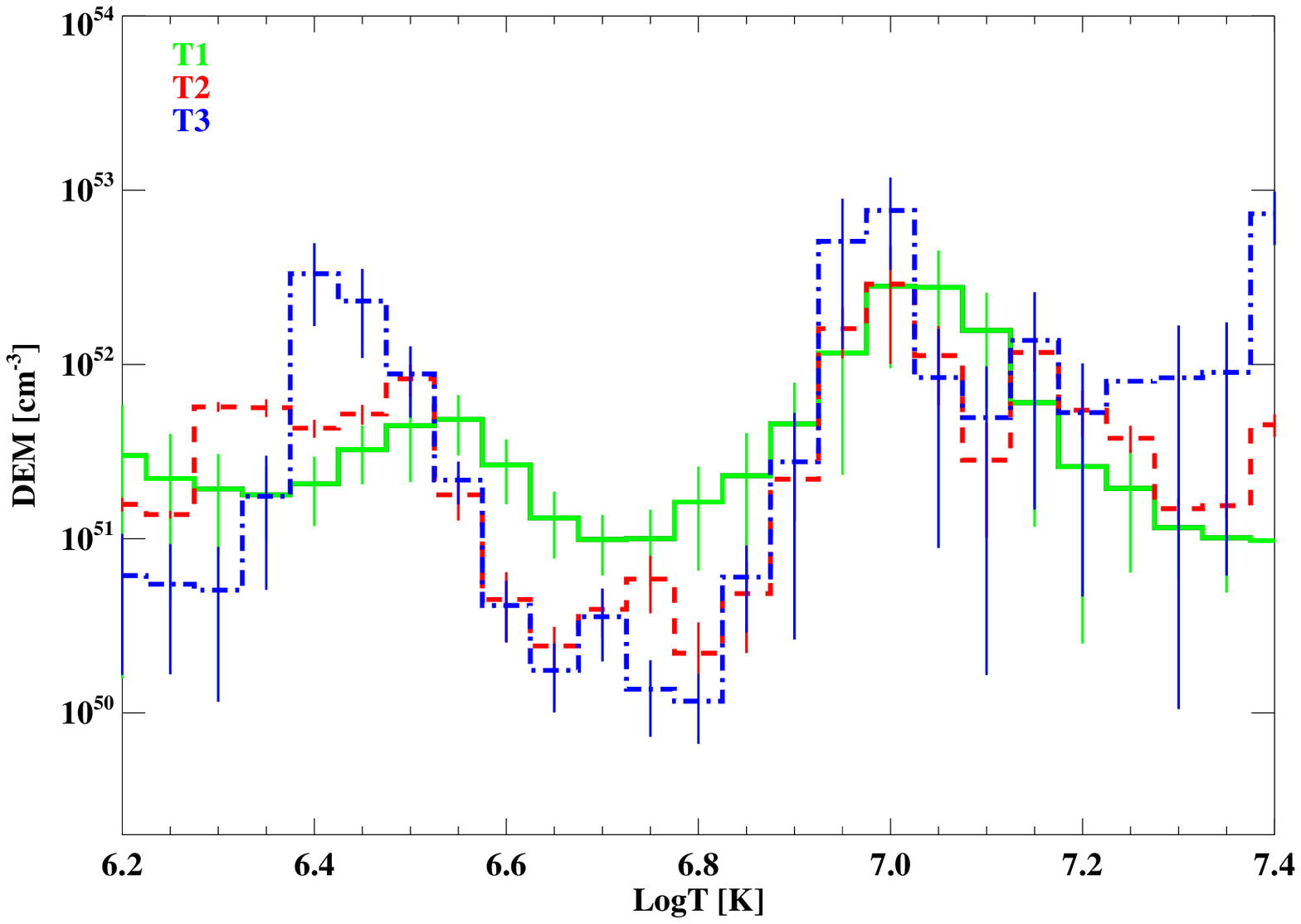}
\caption{\label{fig:dem} Best fit DEM plots versus the log T for HD~155555~AB.}
\end{figure}

In Figure~\ref{fig:dem}, we show the reconstructed DEM for HD~155555~AB.  
The general shape of DEM distribution derived for HD~155555~AB for three time-intervals consistently shows a multi-temperature plasma with a dominant peak at log~T[K]$\sim$6.9-7.0, very well constrained by line fluxes. This peak is consistent with the one of the temperature components predicted by XSPEC model. The peak at the low temperature  between log T[K]$\sim$6.35-6.55 is much more prominent during time interval T3 when compared to the other time intervals. During time interval T2 and T3, we notice a higher temperature component (log T[K]$>$7.35) which implies that the enhanced activity level  during time interval T2 and T3 is adding a discrete source of emission at the higher and lower end of the temperature range.

\begin{figure}[!ht]
\centering
\includegraphics[width=0.45 \textwidth]{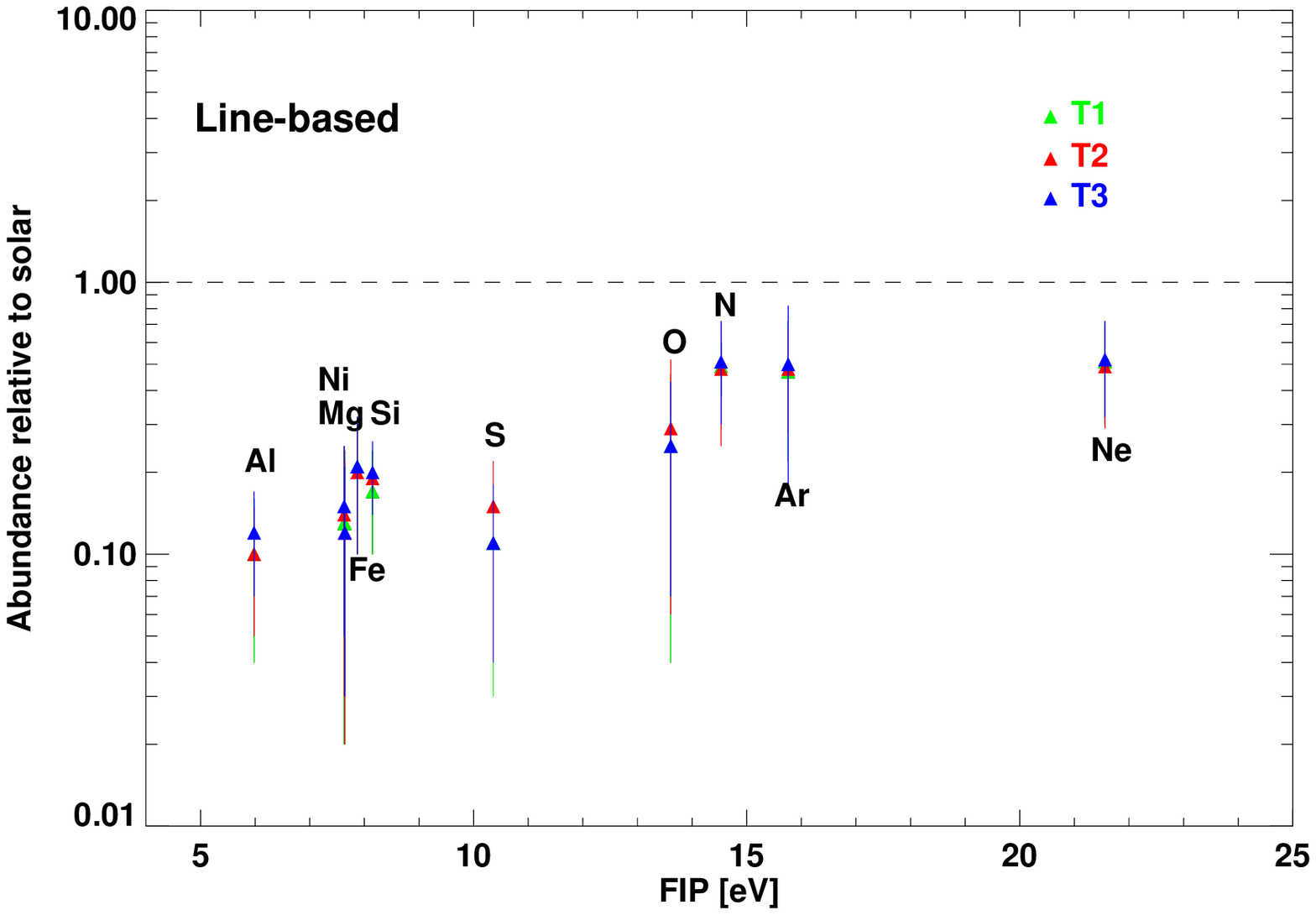}
\includegraphics[width=0.45 \textwidth]{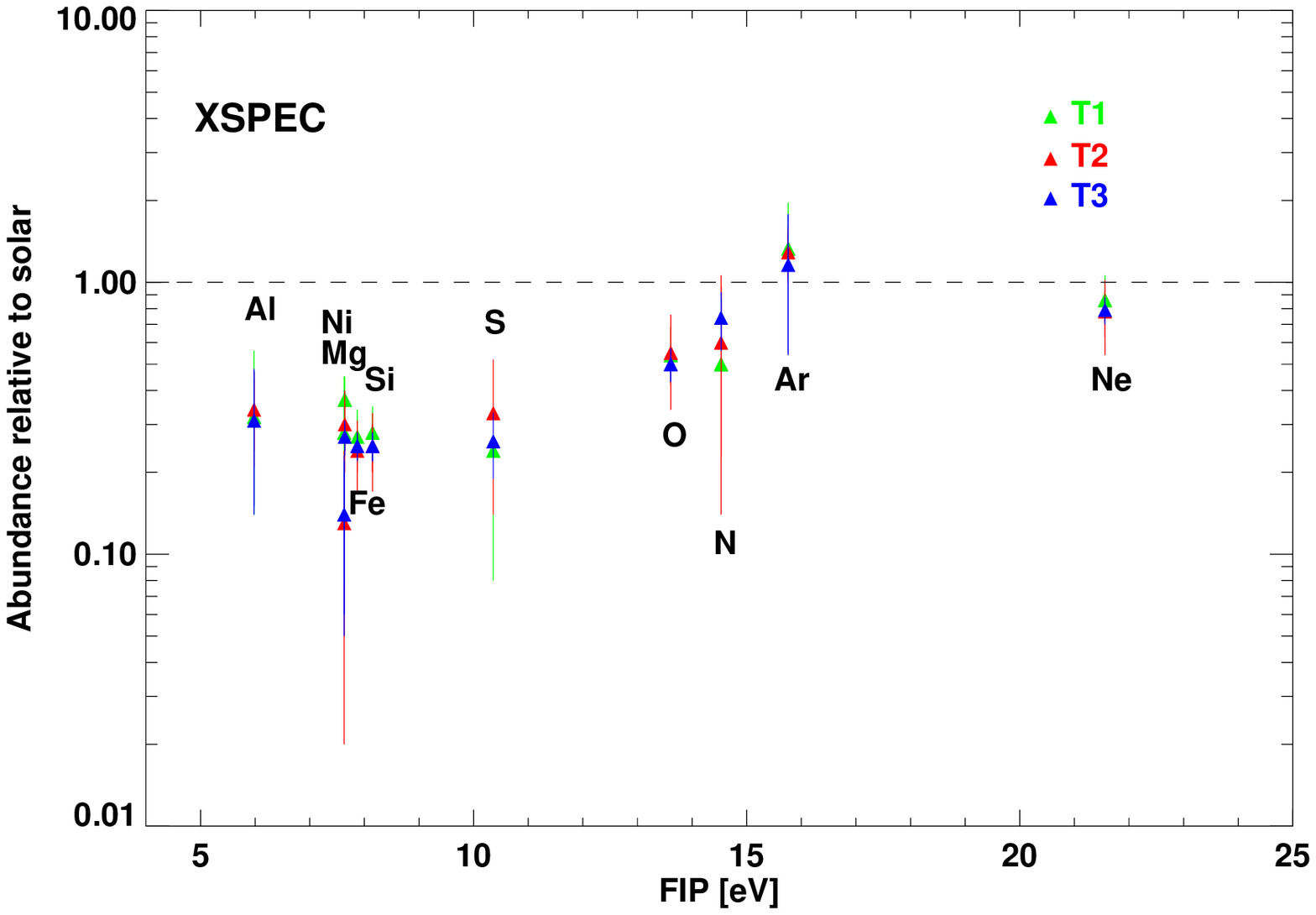}
\caption{\label{fig:fipab}  The coronal abundance relative to solar photospheric values \citep{grevesse_1998} as derived by line-based (left panel) and XSPEC based analysis (right panel) for T1, T2 and T3 time bins plotted as a function of the first ionisation potential (FIP) for HD~155555~AB.}
\end{figure}



\begin{table*}[!ht]
\centering
\caption{\label{tab:abundab}Best fit values of the elemental abundances for different time intervals.} 
\begin{tabular}{ccccccc}
\tableline
\multicolumn{1}{c}{}& \multicolumn{2}{c}{T1}& \multicolumn{2}{c}{T2}& \multicolumn{2}{c}{T3}\\          

Elements & XSPEC & Line-based & XSPEC & Line-based & XSPEC & Line-based\\
\tableline
N &0.50 &0.49$\pm$0.11& 0.60$^{+0.37}_{-0.46}$&0.48$\pm$0.23&0.74$^{+0.18}_{-0.14}$&0.51$\pm$0.21\\
O & 0.54$^{+0.15}_{-0.12}$&0.25$\pm$0.21&0.55$^{+0.21}_{-0.13}$&0.29$\pm$0.23&0.50$^{+0.07}_{-0.07}$&0.25$\pm$0.18 \\
Ne & 0.86$^{+0.20}_{-0.15}$&0.51$\pm$0.21&0.78$^{+0.24}_{-0.15}$&0.49$\pm$0.20&0.79$^{+0.09}_{-0.08}$&0.52$\pm$0.20  \\
Mg & 0.37$^{+0.08}_{-0.07}$ & 0.13$\pm$0.11&0.30$^{+0.10}_{-0.06}$&0.12$\pm$0.10&0.27$^{+0.04}_{-0.03}$&0.12$\pm$0.09 \\
Al & 0.32$^{+0.24}_{-0.18}$&0.10$\pm$0.06&0.34$^{+0.13}_{-0.10}$&0.10$\pm$0.05&0.31$^{+0.16}_{-0.17}$& 0.12$\pm$0.05 \\
Si & 0.28$^{+0.07}_{-0.05}$&0.17$\pm$0.07&0.25$^{+0.08}_{-0.05}$&0.19$\pm$0.02&0.25$^{+0.03}_{-0.03}$& 0.20$\pm$0.06\\
S & 0.24$^{+0.17}_{-0.16}$&0.11$\pm$0.08&0.33$^{+0.19}_{-0.18}$&0.14$\pm$0.07&0.26$^{+0.07}_{-0.07}$& 0.11$\pm$0.07\\
Ar & 1.33$^{+0.63}_{-0.15}$&0.47$\pm$0.25&1.29$^{+0.30}_{-0.25}$&0.48$\pm$0.20& 1.16$^{+0.62}_{-0.61}$&0.50$\pm$0.32 \\
Fe &0.27$^{+0.07}_{-0.05}$ &0.21$\pm$0.11&0.24$^{+0.07}_{-0.04}$&0.20$\pm$0.10&0.25$^{+0.03}_{-0.02}$& 0.21$\pm$0.11 \\
Ni&0.28$^{+0.17}_{-0.20}$&0.13$\pm$0.11&0.13$^{+0.21}_{-0.11}$&0.14$\pm$0.11&0.14$^{+0.09}_{-0.08}$ & 0.15$\pm$0.10\\

\tableline

\end{tabular}

\end{table*}

In Figure~\ref{fig:fipab}, the coronal elemental abundances of HD~155555~AB obtained for all the three time-intervals are plotted as a function of the first ionisation potential of the elements. In Table~\ref{tab:abundab}, we list the coronal elemental abundances relative to the solar photospheric abundances derived by both the line based method and XSPEC fitting  methods. The line-based and the XSPEC fitting to the MEG spectra give nearly consistent values for the elemental abundance except for Ar, despite the difference in the temperature distribution derived using the two methods. This could be because in the line-based analysis we reject lines that are blended and 
problematic whereas the XSPEC fitting assumes that the model can explain each and every individual line in the observed spectrum. Furthermore, we use 3T XSPEC model to characterise the corona which is a criterion we use to define a multi-temperature structure of the corona; whereas the DEM reconstruction shows clearly that the corona of HD~155555~AB has a continuously distributed multi-temperature thermal structure.


\subsubsection{Optical depth effects}
Generally, the X-ray emission from coronal plasmas can be assumed to be optically thin \citep{schrijver_1995, mewe_1995}. In order to justify the optically thin model used in our analysis, we estimated the optical depth by comparing the observed flux ratio of the \ion{Fe}{17} line at 15.265$\textrm{\AA}$ to the resonance line at 15.040$\textrm{\AA}$ of \ion{Fe}{17} but with a lower oscillator strength.  These ratios have been measured in the laboratory using an electron beam ion trap by \cite{brown_1998} and the values range between 0.31-0.36 at 0.85-1.3 keV. The measured value for the Sun is 0.49$\pm$0.05 \citep{saba_1999}. We obtained a ratio of 0.35$\pm$0.11 for HD 155555 AB, which is larger than the value  of 0.26$\pm$0.10 determined for Capella \citep{brinkman_2000} and smaller than the value of 0.49$\pm$0.14 determined for II Peg \citep{huenermoeder_2001}. The uncertainty in the ratio for HD 155555 AB is large enough that the results are consistent with previously estimated ratios.

\subsubsection{Electron densities}

\begin{figure*}[!ht]
\centering
\includegraphics[width=18cm, height=6.5cm]{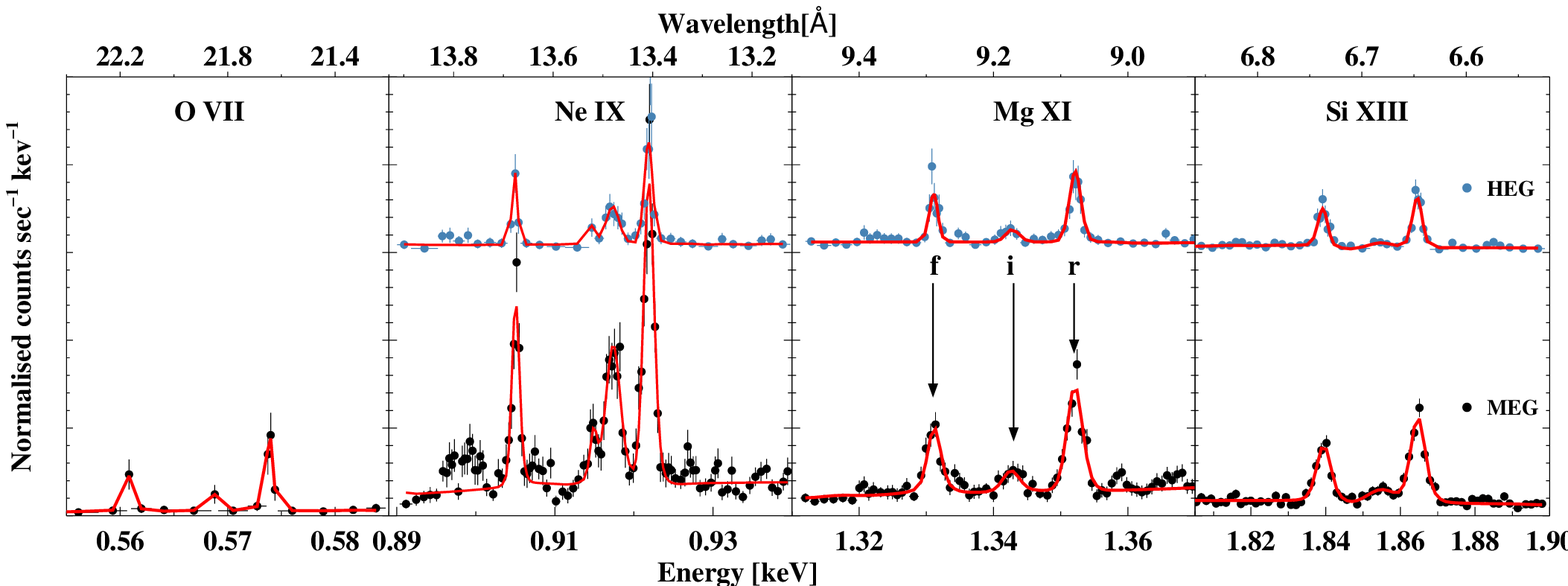}
\caption{\label{fig:fit} He-like triplets of \ion{O}{7}, \ion{Ne}{9}, \ion{Mg}{11} and \ion{Si}{13} measured with the entire HEG and MEG spectra of HD~155555~AB with the line fits. The Gaussian used to fit the lines are shown in red.}

\end{figure*}

We have investigated the electron densities of the coronal plasma using the density-sensitive line flux ratios of forbidden to inter combination lines of helium-like triplets (\ion{N}{6}, \ion{O}{7}, \ion{Ne}{9}, \ion{Mg}{11} and \ion{Si}{13}), based  on the theory of density-sensitive lines as described in detail by \citet{Gabriel_Jordan}. The He-like triplets of \ion{O}{7}, \ion{Ne}{9}, \ion{Mg}{11} and \ion{Si}{13} are strong enough in our observations to obtain characteristic electron densities in the source (see Figure~\ref{fig:fit}). Note that in the second panel of Fig.~\ref{fig:fit}, the \ion{Ne}{9} inter-combination line (13.553 $\textrm{\AA}$) is blended with \ion{Fe}{19} (13.518 $\textrm{\AA}$), since the spectral resolution of \emph{Chandra} HEG and MEG are 0.005 and 0.0025 $\textrm{\AA}$ per bin, respectively, 
the peaks are well-resolved and are easily fitted with two separate Gaussian components. The measured line counts, the f/i ratios and the deduced electron densities for the entire duration of observation are listed in Table~\ref{tab:den}. In order to convert the measured f /i ratios to densities, we approximated the flux ratio by 
 \[
\frac{f}{i} = \frac {R_{o}}{1+\frac{n_{e}}{N_{c}}}
\]
where $R_{o}$ is the low-density limit and $N_{c}$ is the critical density for which we adopted the values from \citet{Pradhan} for each of the ions. Notice that $f/i$ ratio and electron densities obtained from the dispersed spectrum represent mean values over the entire observation. 

%


\begin{table*}[!ht]
\centering
\caption{\label{tab:den} Line fluxes, line ratios and densities for He like triplets of the total (time averaged)  HD~155555~AB spectrum.}
\begin{tabular}{cccccccccc}
\tableline
Ion    & G=$\frac{f+i}{r}$ & log(T$_e$)&R=$\frac{f}{i}$ & log(n$_e$)\\
&&K&&$cm^{-3}$\\
\tableline
\tableline
\ion{Si}{13} & 0.73$\pm$0.12 & 6.92$\pm$0.20&7.08$\pm$1.18 & 13.34$^{+0.83}_{-0.67}$\\
\ion{Mg}{11} & 0.73$\pm$0.07& 6.76$\pm$0.17&1.95$\pm$0.20 &12.72$^{+0.95}_{-0.84}$\\
\ion{Ne}{9} & 0.83$\pm$0.10& 6.48$\pm$0.09 & 2.87$\pm$0.34  & 11.83$^{+1.15}_{-1.08}$ \\
\ion{O}{7}  & 0.82$\pm$0.25 &6.38$\pm$0.22  & 1.13$\pm$0.35  & 10.80$^{+0.81}_{-0.65}$ \\
\tableline
\end{tabular}
\end{table*}

\begin{figure}[!ht]
\centering
\includegraphics[width=9cm, height=6.5cm]{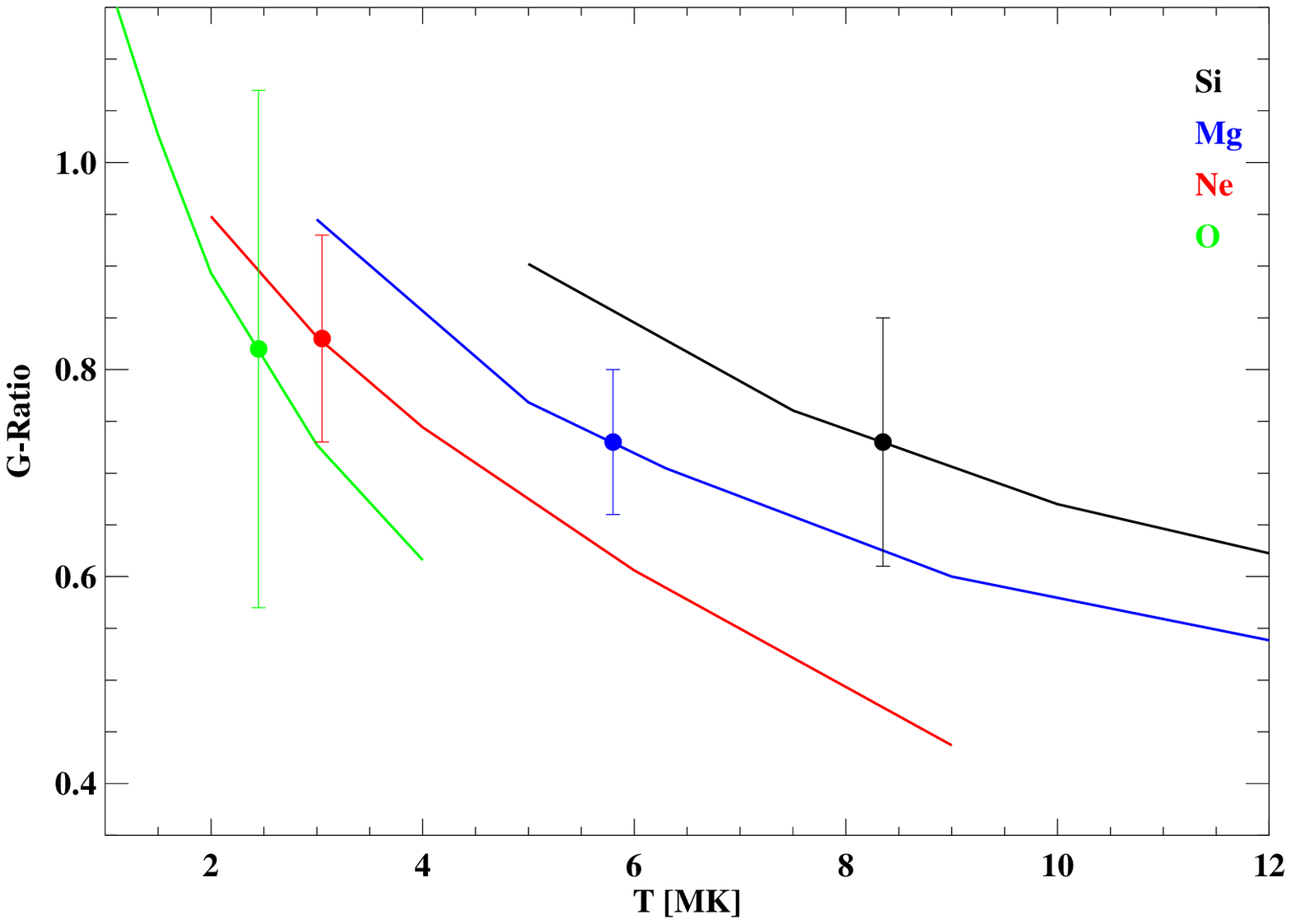}
\caption{\label{fig:grat} The G-ratio curve from \cite{porquet_2000} for \ion{O}{7} (in green), \ion{Ne}{9} (in red), \ion{Mg}{11} (in blue) and \ion{Si}{13} (in black) 
 plotted as a function of temperature. The filled circles represents the measured G-ratio for HD~155555~AB.}

\end{figure}

Furthermore, using the measured f, i and r line fluxes we have calculated the G=$\frac{f+i}{r}$ ratio for the time-averaged spectra, which provides an estimate of the electron temperature T$_e$. The dependence on temperatures of the  G-ratios is due to the collisional excitations which have a different temperature sensitivity for the 
 forbidden and inter-combination lines compared to the r lines. In the case of a strong r line  the G-ratio (G~$\sim$1) suggests  a collisionally dominated plasma (see Table~\ref{tab:den}). We compared the measured G(T) values of \ion{O}{7}, \ion{Ne}{9}, \ion{Mg}{11} and \ion{Si}{13} with the theoretical relation between the electron temperature and G(T) as described by \cite{porquet_2000} and have plotted the results in Figure~\ref{fig:grat}). The comparison shows that the coronal temperature is in 2-12~MK range. This temperature range is comparable to the temperatures derived using the  DEM analysis. 
%

\begin{figure}[!ht]
\centering
\includegraphics[width=9cm,height=13cm, angle=0]{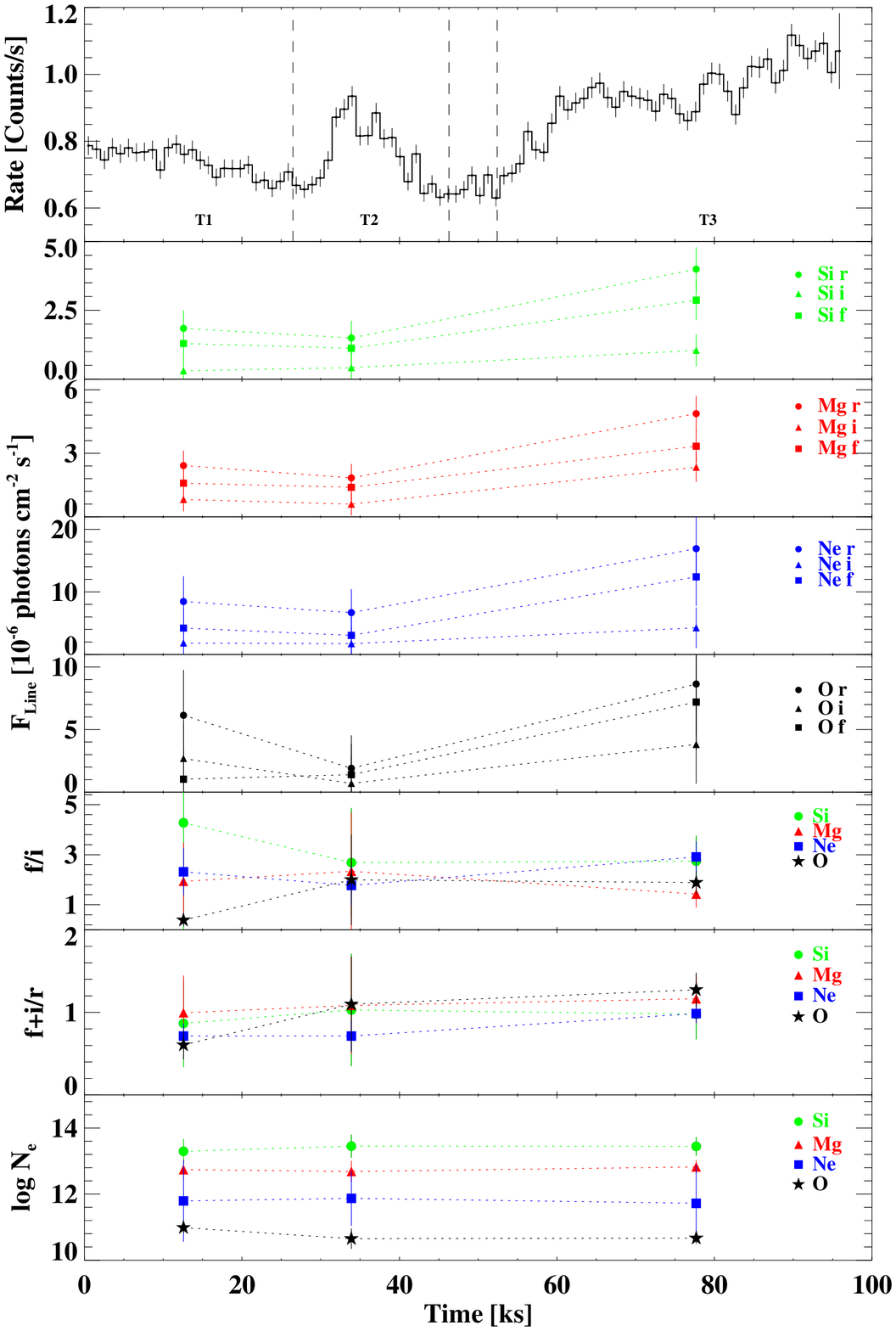}
\caption{\label{fig:lineflux} He-like triplets line fluxes of HD~155555~AB, their ratios and the densities as a function of time.}
\end{figure}

We derived the R and G ratio for each of the time bins and refined the analysis of plasma temperature and density. In Figure~\ref{fig:lineflux}, we show the line fluxes and their ratios as a function of time. Although individual fluxes clearly show an increasing trend during T3, the f/i ratio shows marginal variation between T1, T2 and T3.

The electron densities calculated using \ion{O}{7} triplets for HD~155555~AB time bins T1, T2 and T3 are log~n$_e$[cm$^{-3}$]$\sim10.98^{+0.17}_{-0.98}$, $10.65^{+0.30}_{-0.74}$, and $10.66^{+0.17}_{-0.77}$, respectively at log T[K]$\sim$6.3. The \ion{Ne}{9} triplet is strongly affected by the blending of \ion{Fe}{19} hence the 
obtained density might be highly uncertain; however, we obtain log~n$_e$[cm$^{-3}$]$\sim$11.79$^{+1.23}_{-1.17}$, 11.87$^{+0.83}_{-1.59}$, and 11.72$^{+1.16}_{-0.74}$ for time bins T1, T2 and T3, respectively, at log T[K]$\sim$6.6. Higher densities are indicated by the \ion{Mg}{11} and \ion{Si}{13} triplets yielding log~n$_e$[cm$^{-3}$]$\sim$12.73$^{+1.23}_{-1.17}$, 12.68$^{+0.32}_{-0.67}$, and 12.82$^{+0.23}_{-0.90}$ at log T[K]$\sim$6.9 and log~n$_e$[cm$^{-3}$]$\sim$13.29$^{+0.36}_{-0.52}$, 13.45$^{+0.34}_{-0.58}$, and 13.44$^{+0.27}_{-0.97}$ at log T[K]$\sim$7.0, for time bins T1, T2 and T3, respectively. Our results show that the electron densities are higher in the plasma region with higher temperature.

\section{Analysis and results: HD~155555~C}\label{sec:anal_c}
We subjected the spectra of the M-dwarf companion to the same analysis that we have applied to HD~155555~AB, as described in \S~\ref{sec:anal_ab}. 

\subsection{Light curve}
\begin{figure}[!ht]
\centering
\includegraphics[width=8.5cm, height=7cm]{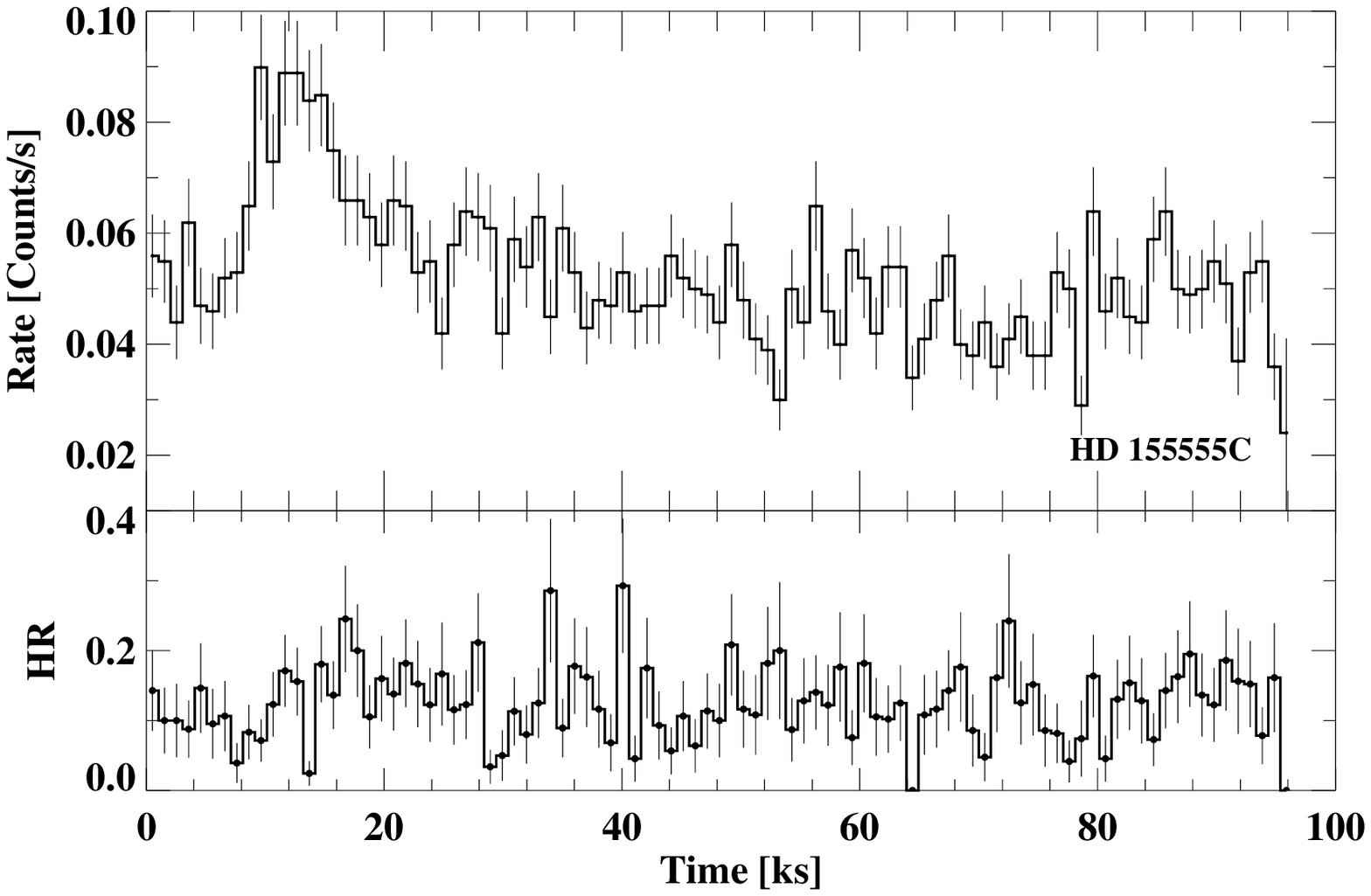}
\caption{\label{fig:lc1} Background subtracted X-ray light curve in 0.4-6.0 keV energy band (top panel) and the hardness ratios as a function of time (bottom panel) for HD~155555~C.} 

\end{figure}

The X-ray light curve obtained in 0.4-6.0 keV range is shown in Figure~\ref{fig:lc1}. There is flare-like feature observed in HD~155555~C where the intensity increases by a factor of $\sim2$ (Fig.~\ref{fig:lc1}). The flare like feature was fitted with an exponential function 
\begin{equation}
CR(t)=A_0 e^{-\frac{(t-t_0)}{\tau_{r,d}}}
\end{equation}
where  CR(t) is the count rate as a function of time, A$_0$ is the count rate at the flare peak, t$_0$ is the time of peak count rate, $\tau_{r,d}$ is the rise/decay time of the flare. The short rise ($\tau_r\sim$6.00$\pm$0.02 ks) followed by a slow steady decline ($\tau_d\sim$8.90$\pm$0.03 ks) in the X-ray counts are  similar to the flares observed on pre-main sequence stars \citep{stelzer_2005, getman_2008}. We have examined the spectral evolution of HD~155555~C but there is no obvious spectral hardening during this flare-like feature (bottom panel of Fig.~\ref{fig:lc1}) based on hardness ratio variation. The enhancement in the emission could occur as a result of magnetic reconnection. Since there is an increase in the count rate and no apparent change in the hardness ratio or temperature, we can interpret this enhancement in two ways 1) the increase in count rate was a result of more plasma with similar coronal properties rotating into the line-of-sight or 2) there is indeed a change in the coronal temperature but could not be observed as a result of the relatively poor statistics of this spectrum.


%

\subsection{Spectral analysis}
The spectrum of the HD~155555~C is significantly softer than that of HD~155555~AB; this is evident clearly in the first-order spectrum (see Fig.~\ref{fig:cspec}). We have used multi-temperature VAPEC model assuming an optically thin coronal plasma to the dispersed spectrum of HD~155555~C. We fit two temperature  component VAPEC models which produced a $\chi^2\sim232$ (DOF$\sim204$). We then added an additional temperature component which produced a $\chi^2\sim164$ (DOF$\sim202$). Based on the F-statistic we obtained a p-value of 0.008. We further added an additional temperature component which did not improve the chi-square significantly and produced an unambiguously large F-test p-value of 0.49. Hence we modelled the spectra of HD~155555~C with three temperature component allowing the abundance of O, Ne, Mg, Si, S and Fe independently. The resultant model parameters are shown in Table~\ref{tab:cspecparam}. We estimate an unabsorbed X-ray flux of $\sim 1.89 \times 10^{-12}$~erg\,cm$^{-2}$\,s$^{-1}$ for HD~155555~C which corresponds to X-ray luminosity of 2$\times10^{29}$ erg s$^{-1}$. The abundances of X-ray emitting plasma obtained from the above model are listed in Table~\ref{tab:cabund}. 


\begin{table}[!ht]
\centering
\caption{\label{tab:cspecparam}  Best-fit spectral parameters obtained from iterative fitting of dispersed X-ray spectra for HD~155555~C data using 3T VAPEC plasma models.}
\begin{tabular}{lcccccccccc}
\tableline
Parameters &\\
\tableline
log t$_1$ drive
drive[K] & 6.49$^{+0.12}_{-0.20}$\\
EM$_1$ [cm$^{-3}$] & 51.65$^{+0.02}_{-0.03}$\\
log t$_2$ [K] & 6.90$^{+0.05}_{-0.12}$\\
EM$_2$ [cm$^{-3}$] & 51.71$^{+0.08}_{-0.09}$ \\
log t$_3$ [K] & 7.34$^{+0.05}_{-0.04}$\\
EM$_3$ [cm$^{-3}$] & 51.86$^{+0.09}_{-0.10}$ \\
red. $\chi^2$& 0.81 \\
DOF &202\\
log~$\mathrm{L}_\mathrm{x}$ [erg\,s$^{-1}$]&29.30\\
\tableline
\end{tabular}

\footnotesize{Note: the errors are estimated with 90\% confidence limit.}\\
\end{table}

\begin{figure}[!ht]
\centering
\includegraphics[width=0.45 \textwidth]{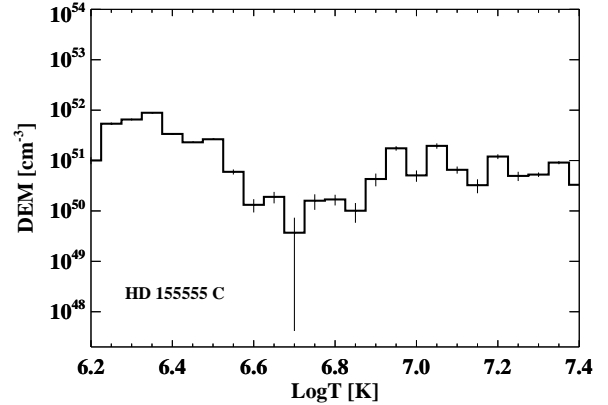}
\caption{\label{fig:cdem} Best fit DEM plots versus the log T for  HD~155555~C.}

\end{figure}

In addition to the global fits we have used the line-based analysis as in the case of HD~155555~AB to assess the coronal temperatures, emission measures and the abundances of HD~155555~C. We reconstructed the DEM using all the strong lines. The resulting DEM distribution is shown in Figure~\ref{fig:cdem} peaking at log~T[K]$\sim$6.3. The elemental abundances were also determined using the reconstructed DEM and their values are given in Table~\ref{tab:cabund}.

\begin{table}
\centering
\caption{\label{tab:cabund} Abundances measurements for HD 155555 C.}
\begin{tabular}{cccc}
\tableline
\multicolumn{1}{c}{}&\multicolumn{2}{c}{HD~155555~C}\\          

Elements & XSPEC & Line-based\\
\tableline
O & 0.31$_{-0.18}^{+0.39}$&0.29$\pm$0.06\\
Ne & 0.92$_{-0.90}^{+1.03}$&0.89$\pm$0.45\\
Mg & 0.36$_{-0.06}^{+0.17}$&0.27$\pm$0.15\\
Al & 0.5 &0.45$\pm$0.31\\
Si &  0.48$_{-0.22}^{+0.28}$&0.29$\pm$0.10\\
S & 0.55$^{+0.30}_{-0.24}$&0.35$\pm$0.12\\
Fe & 0.50$_{-0.20}^{+0.33}$&0.41$\pm$0.14\\

\tableline

\end{tabular}

\end{table}

\begin{figure}[!ht]
\centering
\includegraphics[width=0.45 \textwidth]{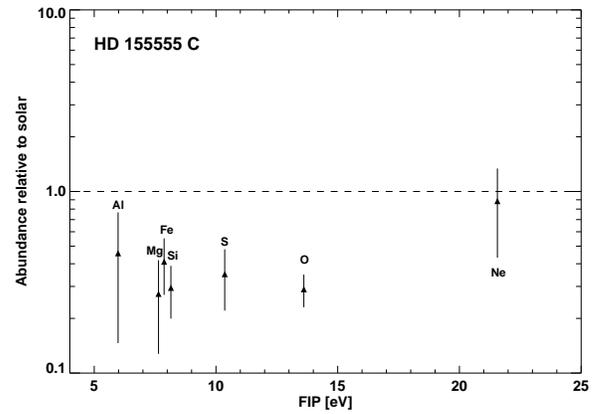}
\caption{\label{fig:fipc}  The coronal abundance relative to the solar photospheric values \citep{grevesse_1998} as derived by line-based analysis are plotted as a function of the first ionisation potential (FIP) for HD~155555~C.}
\end{figure}


 We estimated the coronal density 
using density-sensitive ratio of forbidden and inter-combination 
lines of \ion{Ne}{9} triplet and obtained a value of 
log~N$_e$$\sim$11.78$^{+0.57}_{-0.23}$ (Table~\ref{tab:cden}). 
Using the line fluxes of \ion{Si}{13}, \ion{Mg}{11} and \ion{O}{7} 
we estimate an upper limit to the coronal densities of these elements  (Table~\ref{tab:cden}).

\begin{table*}
\centering
\caption{\label{tab:cden} Line fluxes, line ratios and density for He like triplets of HD~155555~C.}
\begin{tabular}{cccccccccc}
\tableline
Ion    &f & i & r & G=$\frac{f+i}{r}$ & log(T$_e$)&R=$\frac{f}{i}$ & log(n$_e$)\\
&[counts]&[counts]&[counts]&&K&&$cm^{-3}$\\
\tableline

\tableline
\ion{Si}{13} & 9.63$\pm$4.22&$<$3.27&28.23$\pm$6.38&$<$0.45&...&$<$2.94&$<$13.42\\
\ion{Mg}{11} & 26.30$\pm$6.20 &$<$8.49& 15.59$\pm$5.04&$<$2.23&...&$<$3.09&$<$12.59\\
\ion{Ne}{9} & 22.28$\pm$5.79&9.25$\pm$4.16&13.63$\pm$4.79&2.31$\pm$1.14&...&2.40$\pm$1.25&11.78$^{+0.57}_{-0.23}$\\
\ion{O}{7} & 4.48$\pm$3.28&$<$2.43&$<$5.49&$<$1.25&...&$<$1.84&$<$10.67\\

\tableline

\end{tabular}

\end{table*}


The ratio of \ion{Fe}{17} resonance line at 15.01\AA ~and the adjacent line at 
15.26~\AA~yielded 0.38$\pm$0.25, indicating a ratio similar to HD~155555~AB.


\section{Discussion}\label{sec:disc}

Our observation of HD~155555~AB shows continuous variability such that no part of the observation can be defined as being ''quiescent'' (see Fig.\ref{fig:lc}). The light curve was divided into three different time intervals. Interval T1 shows small-scale fluctuations and a steady decline which is perhaps indicative of the rotational modulation. During interval T2, a flare-like event is observed. The flare intensity is correlated with the hardness ratio. During interval T3, a steady rise with several small-scale variations is seen. This could be the result of one of the binary components crossing a large active region or several such regions on the stellar surface.

Our analysis of the \emph{Chandra} HETG spectra has provided an insight into the plasma emission measure distribution, coronal temperatures, the abundances of individual elements, and the plasma densities as determined by He-like triplets. An obvious feature of the reconstructed DEMs (Fig.\ref{fig:dem}) for each time interval is the apparent bimodal structure at similar temperatures. By comparing the XSPEC models predicting the observed spectra during three different time bins we find an increase in the source luminosity from 3.2$\times10^{30}$ erg s$^{-1}$ (during T1) to 4.1 $\times10^{30}$ erg s$^{-1}$ (during T3). This is also reflected in the DEMs where the high-temperature peak at log T[K]$\sim$6.9-7.0 is higher during T3 by a factor 1.5-2 when compared to that during T1. The high-temperature peak values of the DEM are consistent with the values obtained using XSPEC. The peaks of the low-temperature components of the DEM (at log T[K]$\sim$6.3-6.5) increase by factors of 2-10 for T2 and T3 compared to T1. The large emission measure can be a result of an increase in the density or increase in the emitting volume.

\begin{figure}[!ht]
\centering
\includegraphics[width=0.45 \textwidth]{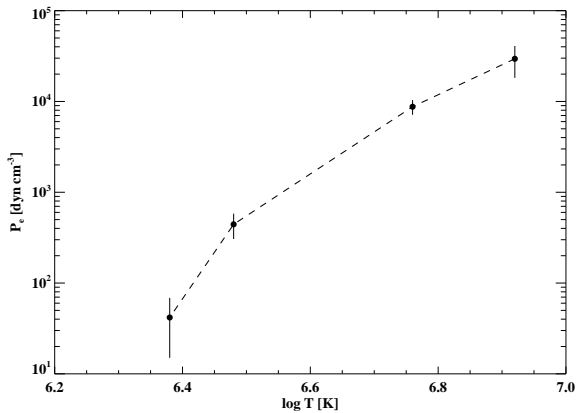}
\caption{\label{fig:pressure} Electron pressure derived from the measured electron densities at different temperatures.}
\end{figure}

A key parameter to interpret magnetically confined coronal structure is the plasma density. The measurements of the density and temperature obtained from the detailed analysis of He-like triplets allow us to estimate the volume and thus the average pressure of the emitting coronal region where these He-like triplets are formed. In Figure~\ref{fig:pressure}, we plot the estimated plasma pressure obtained using the densities and temperatures derived from the triplet line ratios as a function of the temperature. The plasma pressure increases steeply with increase in the temperature. A similar trend has been observed previously by \cite{argiroffi_2003} and \cite{sanz_2003}; they interpret this due to the presence of different kinds of coronal structures in different isobaric conditions where the plasma pressure increases for increasing temperature of the confined plasma but the volume filling factor decreases.

Stellar coronae often exhibit abundance anomaly patterns. A few moderately active, and even some low activity stars like the Sun, show an enhancement of low FIP elements and a depletion of high FIP elements. Highly active stars, on the other hand, exhibit an inverse pattern with the depletion of low-FIP elements and enhancement of high-FIP elements (inverse-FIP or iFIP effect). These effects result in higher metallicities in the coronae of stars with low activity compared to high activity stars \citep{guedel_2001}. The abundances derived for HD~155555~AB by the two methods used here (see \S\ref{sec:xspecab} and \S\ref{sec:lineab}) show a general sub-solar abundance for most of the elements, though the low FIP elements ($<$10 eV; e.g. Mg, Si, Fe) are depleted more compared to the high FIP elements ($>10$ eV; e.g. O, Ne, Ar, N) (Fig.~\ref{fig:fipab}). The coronal abundances of HD~155555~AB relative to solar photospheric values \citep{grevesse_1998}, do not show a strong FIP dependence, but rather suggest a moderate iFIP effect. We do not see any strong variation in the individual abundances during different time intervals in HD~155555~AB, indicating that there is no significantly fresh chromospheric material being evaporated into the corona.

\subsection{Comparison with other stars}
In order to check whether there are any differences between the coronae  of
stars that are active due to their extreme youth and those in RS CVn binary systems that have been spun up by tidal forces, we compare the spectrum of HD 155555 AB with that of active stars and young stars, viz., 
AR Lac (an active RS Cvn type binary), AB Dor (an active PMS star), HD 98800 (a WTTS), and TW Hya (a CTTS). AR Lac is one of the brightest RS CVn type binary \citep{siviero_2006, drake_2014} with an orbital period of 1.98 d and with its binary components separated by $\sim$9.2 R$_{\odot}$ \citep{chambliss_1976,popper_1977}. 
AB Dor is a bright (V$\approx$6.9) young (age$\sim$30-100 Myr) ultra-fast rotating (P$_{rot}\sim0.52$d) late-type star (\citealt{amado_2001, guirado_2011} 
and references therein), and a member of a wide visual binary system (its companion is Rst 137B at a separation of 9$''$ with spectral type M3-M5; \citealt{guirado_2011}). AB Dor is a highly active star that has been observed by many space-based observatories across the UV, EUV and X-ray wavelengths \citep{sanz_2003, garcia_2008, lalitha_2013b}. TW Hya is a low-mass  K-type star with an age of $<$10 Myr. TW Hya is accreting from its circumstellar disk  (accretion rate of $\sim$1.0$\times10^{-9}$ and 5$\times10^{-8}$ M$_{\odot}$ yr$^{-1}$; \citealt{batalha_2002, brickhouse_2010}), which may give rise to post-shock plasma. Therefore, we have also chosen a non-accreting X-ray bright PMS star, HD 98800 for comparison. TW Hya is a single CTTS, whereas HD 98800 is a multiple system with two visual components separated by 0.8$''$ and orbital period of 300-400 years \citep{soderblom_1998, tokovinin_1999, prato_2001, boden_2005}. The two components of HD 98800 are well resolved in the \emph{Chandra} HETG zeroth-order image with the primary component HD 98800 A appearing to be $\sim$ 4 times brighter than HD 98800 B \citep{kastner_2004}. This system displays an X-ray luminosity similar to TW Hya, however, there is no evidence of accretion in HD 98800 \citep{kastner_1997, kastner_2004}.

\begin{figure}[!ht]
\centering
\includegraphics[width=9cm,height=9.5cm, angle=0]{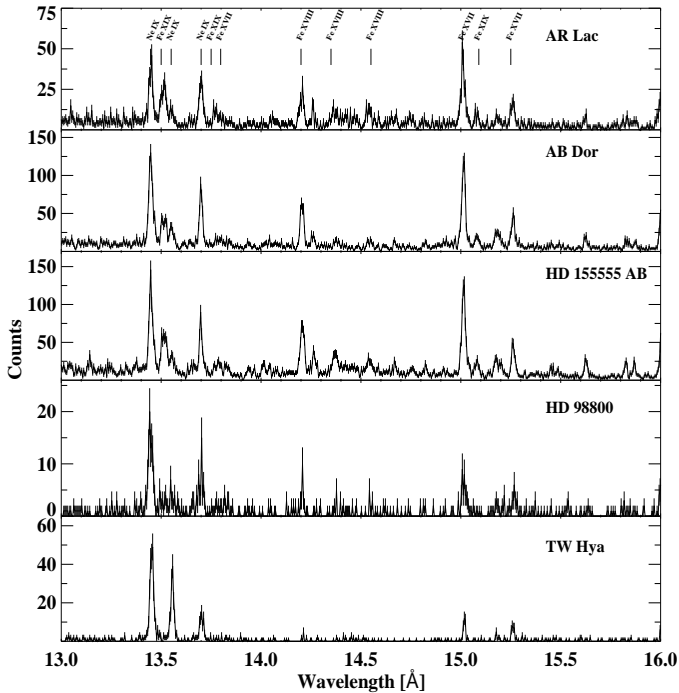}
\caption{\label{fig:compab} \emph{Chandra} HETG spectra of the active late-type 
star AR~Lac, young active stars AB Dor, HD~155555~AB, the pre-main sequence stars HD 98800 (WTTS) and TW~Hya (CTTS), in the spectral region encompassing of \ion{Ne}{9} He-like triplets and \ion{Fe}{17} lines.}

\end{figure}

The basic stellar and X-ray properties of these stars are summarized in Table~\ref{tab:comp}. The coronal emission of AR Lac, TW Hya and HD 98800 has been studied in detail via \emph{Chandra} HETG observations \citep{batalha_2002, huenemoerder_2004, kastner_2004, drake_2014}. In Figure~\ref{fig:compab}, we show the \emph{Chandra} HETG spectra in the 13-16 $\textrm{\AA}$ region  for AR Lac, AB Dor, HD 155555 AB, TW Hya and HD 98800. This wavelength region encompasses several highly ionised Fe lines and the \ion{Ne}{9} triplets. Visual comparison shows that the spectrum of HD~155555~AB is very similar to that of AR Lac and AB Dor showing similar relative line strengths for the \ion{Ne}{9} triplet and the Fe lines. On the other hand, TW Hya and HD 98800 show the \ion{Ne}{9} resonance line to be strong, and Fe lines to be weak compared to HD~155555~AB, AR Lac, and AB Dor.

\begin{table*}[!ht]
\footnotesize
\centering
\caption{\label{tab:comp} Comparison of physical properties of AR Lac, AB Dor, HD~155555~AB, HD 98800 and TW Hya.}
\begin{tabular}{lccccccccccc}
\tableline
Star    & Spectral & Age & Binarity & Ongoing  & P$_{rot/orb}$ & Distance & log~$\mathrm{L}_\mathrm{x}$& log~$\frac{\mathrm{L}_\mathrm{x}}{\mathrm{L}_\mathrm{bol}}$& T \\
        &Type& Myr  &			 & Accretion & days  &		&	  &							 & MK\\
\tableline
\tableline
 AR Lac       &K0IV+G5IV   & $>$3000$^1$ & Y  & N & 1.98 & $\sim$43 &30.89 & -3.45 &$\sim$16\\
 AB Dor       &K0-2V       & 30-100  & Y  & N & 0.52 & $\sim$15 &30.01 & -3.00 &$\sim$ 13\\
 HD~155555~AB &G5IV+K0IV & 18 	  & Y  & N & 1.68 & $\sim$31&30.60 & -3.28 &$\sim$ 12\\
HD 98800      &K4V         & 10 	  & Y  & N & 262 &	$\sim$40& 29.83 & -3.40 & 2.5-10 \\
TW Hya        &K6Ve        & 8 	  & N  & Y & 4.5? &$\sim$57 &30.11 & -2.92	& $\sim$3\\
\tableline
\end{tabular}

\footnotesize{Note: The ages, spectral types and distances are from 1) \citealt{lanza_1998, drake_2014}, and references therein, for AR Lac; 2) \citealt{guirado_2011} and references therein for AB Dor; 3) \citealt{strassmeier_2000} for HD 155555 AB; 4) \citealt{kastner_1997, webb_1999} for HD 98800; 5) \citealt{batalha_2002,setiawan_2008} for TW Hya. For AB Dor, P$_{rot/orb}$ listed in column 6 represents the rotation period \citep{guirado_2011}, whereas, for HD 98800 this value represents the orbital period of the wide binary \citep{boden_2005}. For AR Lac and HD 155555~AB, P$_{rot/orb}$ represents both rotation period and the orbital period since these systems are tidally locked \citep{lanza_1998,strassmeier_2000}. For TW Hya, P$_{rot/orb}$ listed in column 6 represents an uncertain rotation period \citep{alencar_2002}. The coronal temperature (T) are the emission measure weighted temperatures in MK. For AR Lac and AB Dor the temperature and emission measures are obtained from \cite{singh_1996a} and \cite{lalitha_2013}, respectively. For TW Hya and HD 98800, the temperatures are obtained from \cite{kastner_2002} and \cite{kastner_2004}, respectively.}\\
\end{table*}


%

Comparison of our results with AR Lac, AB Dor, TW Hya, and HD 98800 indicate that HD~155555~AB is very similar to AR Lac and AB Dor in its emission measure distribution, abundances, and electron densities, despite the difference in their ages. During flare-like events the emission measures peak at similar temperatures (log T[K]$\sim$6.3 and 6.9) for HD~155555~AB, AR Lac \citep{huenermode_2002} and AB Dor \citep{sanz_2003}. However, an additional higher temperature component are observed during both quiescence and flare at log T[K] $\sim$ for AB Dor. This difference in the temperatures could be due to the rapid rotation of AB Dor \citep{randich_1998} or due to its higher activity level \citep{telleschi_2005}. We note that the DEM distributions for both AR Lac and AB Dor are trimodal, whereas this is the case only during flare-like events (T2) for HD 155555~AB. 

The total emission measure for the higher temperature component (log T [K]$\sim$6.9) is $\sim$52.35$^{+0.08}_{-0.06}$ cm$^{-3}$ for HD 155555 AB, cf., $\sim$52.95$^{+0.05}_{-0.06}$ and $\sim$52.34$^{+0.03}_{-0.07}$ cm$^{-3}$ for AR Lac and AB Dor, respectively (\citealt{huenermode_2002,sanz_2003}). For TW Hya and HD 98800, there is no  emission component at log T [K]$\sim$6.9. The DEM for TW Hya peaks at log T[K]$\sim$6.5 \citep{kastner_2002} with no evidence of a hotter component (log T[K]$>$7.0). Similarly the DEM for HD 98800 has one peak at log T[K]$\sim$6.5, and a flat DEM thereafter \citep{huenemoerder_2004, kastner_2004}.

\cite{guedel_1997} studied a sample of solar-like G stars with ages ranging from 70-9000 Myr and found that the emission  measure of the hotter component rapidly decreases with age. Our study, however, shows that the emission measures of the hotter component for HD 155555 AB and AB Dor are very similar with no noticeable trend. We note that the age of HD 155555 AB is several Myr younger than the youngest star (EK Dra, age$\sim$ 70 Myr) in the sample compiled by \cite{guedel_1997}. A more detailed analysis and interpretation of the coronal temperatures, emission measures, and densities of K-type stars with a range of ages is required in order to characterise how the X-ray properties of these stars evolve with time, and how  their behaviour compares with solar-type stars.  

The ratio of Ne to Fe abundance is  $\sim$3 for HD~155555~AB , $\sim$5 for AR Lac \citep{huenermode_2002},  $\sim$5 for AB Dor \citep{sanz_2003} and $\sim$5 for HD 98800 \citep{kastner_2004}, whereas this ratio is larger than the solar value by a factor of 10 for TW Hya \citep{kastner_2002}. Although the DEM peaks are similar for TW Hya and HD 98800, the Ne/Fe abundance ratio for HD 98800 is unlike that seen in TW Hya. This large ratio in TW Hya could be the result of an ongoing accretion where significant fractionation occurs in the circumstellar disk.
 
The electron densities (log n$_e$) obtained for HD~155555~AB with \ion{O}{7} and \ion{Ne}{9} are 10.80 and 11.83 cm$^{-3}$, respectively, which are similar to the values for AR~Lac \citep{huenermode_2002} and AB Dor \citep{sanz_2003,lalitha_2013}. However, the log n$_e$ values for \ion{O}{7} and \ion{Ne}{9} in TW Hya are 11.75 and 12.47 cm$^{-3}$ respectively \citep{kastner_2002}. This could be the result of ongoing accretion seen in TW~Hya \citep{argiroffi_2009, brickhouse_2010, dupree_2012, ness_2005}.

\begin{figure}[!ht]
\centering
\includegraphics[width=9cm,height=9.5cm, angle=0]{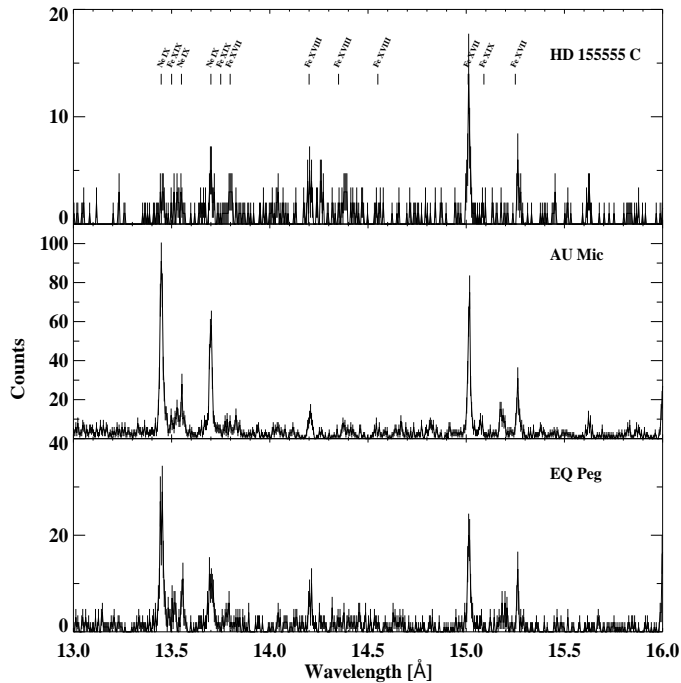}
\caption{\label{fig:compc} \emph{Chandra} HETG spectra of HD~155555~C and the active late-type M stars AU Mic and EQ Peg with similar spectral type as HD~155555~C system, 
in the spectral region encompassing of \ion{Ne}{9} He-like triplets and \ion{Fe}{17} lines.}

\end{figure}

Similarly, we have compared our results on HD 155555~C with other studies of M stars. In Figure~\ref{fig:compc}, we show the \emph{Chandra} HETG spectra in the 13-16 $\textrm{\AA}$ region for HD~155555~C, AU Mic and EQ Peg. AU Mic is an active nearby (distance$\sim$10 pc)  M type star showing typical X-ray luminosity of 2$\times$10$^{29}$ ergs s$^{-1}$, similar to that of HD~155555~C \citep{rodono_1986,del_2002}. EQ Peg is a visual binary consisting of two stars with spectral type M3.5 and M4.5 which are separated by $\approx$30 AU. At a distance of 6.1 pc, the quiescent X-ray luminosity of EQ Peg has been measured to be between 6-7$\times$10$^{28}$ ergs s$^{-1}$ \citep{robrade_2005}. EQ Peg has a spectral type similar to that of HD~155555~C, and AU Mic shows X-ray emission similar to that of HD~155555~C. Visual comparison of Figure~\ref{fig:compc}  shows that the spectrum of HD~155555~C shows a strong flux at 15.01 $\textrm{\AA}$ similar to what is seen in EQ Peg and AU Mic. The \ion{Ne}{9} triplet emission which is very strong in both EQ Peg and AU Mic is barely visible in the case of HD~155555~C. Although the \ion{Ne}{9} line luminosity for HD 155555~C is at least half an order of magnitude brighter than that of AU Mic, the Ne/Fe abundance ratio is $\sim$2, whereas for AU Mic and EQ Peg the ratios are $\sim$ 7 and $\sim$8, respectively.

%
The emission measure of HD~155555~C shows a peak at log~T[K]$\sim$6.4, and has no significant plasma emission at log T[K]$\sim$7. In contrast, AU Mic and EQ Peg show peaks around log T[K]$\sim$6.8-6.9 with the bulk of the plasma located between  log T[K]$\sim$6.3-7.3. However, the temperatures and emission measures obtained by XSPEC model fit are consistent with the results obtained for AU Mic and EQ Peg. The abundances of HD~155555~C show a moderate iFIP effect, which is common among stars with high activity levels like AU Mic and EQ Peg. The metallicities of HD~155555~C found here are sub-solar. 

\section{Conclusion}\label{sec:sum}

We have analysed the \emph{Chandra} HETG data of a young active multiple system, HD 155555. We infer the following characteristics of the coronal plasma of this system: 

\begin{itemize}
\item The X-ray emission shows continuous variability throughout the observation, with flare-like events. The light  curve also shows a possible modulation due to coronal hot spots. Since the observation reported here spans only $\sim65\%$ (P$_{rot}\sim1.68$d and P$_{obs}\sim1.09$d) of the rotation period, additional observations covering at least a few rotation periods of HD 155555 AB are required to verify whether the observed variation is indeed due to rotational modulation. 
\item The emission measure distributions calculated for the coronal plasma of HD~155555~AB and HD~155555~C using individual line fluxes, show a stable structure at  $\log(T)=6.4$ followed by a steep increase up to the peak at  $\log(T)=6.9$. During high activity state the DEM shows an increase in the amount of X-ray emission. For HD~155555~C, the DEM shows a peak at  $\log(T)=6.3$, followed by a plateau in the range $\log(T)=6.9$-7.4.
\item Element abundances in the corona of HD~155555~AB and HD~155555~C are observed to follow an intermediate behavior between the solar-like FIP effect, and the inverse FIP effect observed in other active stars. The coronal metallicity in both the HD 155555 AB and C spectra is sub-solar, as is typically found in active stars.
\item The continuum, line emissivity, densities and the temperatures of the coronal plasma in HD~155555~AB, are very similar to that of other active stars of similar age, like AB Dor and older RS CVns, like AR Lac, but very different from that of other young stars of CTTS or WTTS type. These results thus indicate that age has little effect on their activity, unlike what is seen in the solar-type stars. A detailed study of a range of K-type stars with different ages, however, may reveal the dependence of coronal properties on age, and whether the behaviour is similar to the solar-type stars or not. 
\item The X-ray emission of the HD 155555 C component is similar to that of other M-type dwarf stars of similar activity levels in the solar neighborhood.
\end{itemize}

\bibliographystyle{apj}
\bibliography{paper}

\begin{thebibliography}{}

\bibitem[\protect\citeauthoryear{{Alencar} \& {Batalha}}{{Alencar} \&
  {Batalha}}{2002}]{alencar_2002}
{Alencar}, S.~H.~P.,  \& {Batalha}, C. 2002, \apj, 571, 378

\bibitem[\protect\citeauthoryear{{Amado} et~al.}{{Amado}
  et~al.}{2001}]{amado_2001}
{Amado}, P.~J., {Cutispoto}, G., {Lanza}, A.~F.,  \& {Rodon{\`o}}, M. 2001, in
  Astronomical Society of the Pacific Conference Series, Vol. 223, 11th
  Cambridge Workshop on Cool Stars, Stellar Systems and the Sun, ed. R.~J.
  {Garcia Lopez}, R.~{Rebolo}, \& M.~R. {Zapaterio Osorio}, 895

\bibitem[\protect\citeauthoryear{{Argiroffi}, {Maggio}, \& {Peres}}{{Argiroffi}
  et~al.}{2003}]{argiroffi_2003}
{Argiroffi}, C., {Maggio}, A.,  \& {Peres}, G. 2003, \aap, 404, 1033

\bibitem[\protect\citeauthoryear{{Argiroffi} et~al.}{{Argiroffi}
  et~al.}{2009}]{argiroffi_2009}
{Argiroffi}, C., {Maggio}, A., {Peres}, G., {Drake}, J.~J.,
  {L{\'o}pez-Santiago}, J., {Sciortino}, S.,  \& {Stelzer}, B. 2009, \aap, 507,
  939

\bibitem[\protect\citeauthoryear{{Arnaud}}{{Arnaud}}{1996}]{arnaud_1996}
{Arnaud}, K.~A. 1996, in ASP Conf. Ser. 101: Astronomical Data Analysis
  Software and Systems V, ed. G.~H. {Jacoby} \& J.~{Barnes}, 17

\bibitem[\protect\citeauthoryear{{Audard} et~al.}{{Audard}
  et~al.}{2002}]{audard_2002}
{Audard}, M., {G{\"u}del}, M., {Sres}, A., {Mewe}, R., {Raassen}, A.~J.~J.,
  {Behar}, E., {Foley}, C.~R.,  \& {van der Meer}, R.~L.~J. 2002, in
  Astronomical Society of the Pacific Conference Series, Vol. 277, Stellar
  Coronae in the Chandra and XMM-NEWTON Era, ed. F.~{Favata} \& J.~J. {Drake},
  65

\bibitem[\protect\citeauthoryear{{Audard} et~al.}{{Audard}
  et~al.}{2003}]{audard_2003}
{Audard}, M., {G{\"u}del}, M., {Sres}, A., {Raassen}, A.~J.~J.,  \& {Mewe}, R.
  2003, \aap, 398, 1137

\bibitem[\protect\citeauthoryear{{Ball} et~al.}{{Ball}
  et~al.}{2005}]{ball_2005}
{Ball}, B., {Drake}, J.~J., {Lin}, L., {Kashyap}, V., {Laming}, J.~M.,  \&
  {Garc{\'{\i}}a-Alvarez}, D. 2005, \apj, 634, 1336

\bibitem[\protect\citeauthoryear{{Barstow}}{{Barstow}}{1987}]{barstow_1987}
{Barstow}, M.~A. 1987, \mnras, 228, 251

\bibitem[\protect\citeauthoryear{{Batalha} et~al.}{{Batalha}
  et~al.}{2002}]{batalha_2002}
{Batalha}, C., {Batalha}, N.~M., {Alencar}, S.~H.~P., {Lopes}, D.~F.,  \&
  {Duarte}, E.~S. 2002, \apj, 580, 343

\bibitem[\protect\citeauthoryear{{Boden} et~al.}{{Boden}
  et~al.}{2005}]{boden_2005}
{Boden}, A.~F., et~al. 2005, \apj, 635, 442

\bibitem[\protect\citeauthoryear{{Brickhouse} et~al.}{{Brickhouse}
  et~al.}{2010}]{brickhouse_2010}
{Brickhouse}, N.~S., {Cranmer}, S.~R., {Dupree}, A.~K., {Luna}, G.~J.~M.,  \&
  {Wolk}, S. 2010, \apj, 710, 1835

\bibitem[\protect\citeauthoryear{{Brinkman} et~al.}{{Brinkman}
  et~al.}{2001}]{brinkman_2001}
{Brinkman}, A.~C., et~al. 2001, \aap, 365, L324

\bibitem[\protect\citeauthoryear{{Brinkman} et~al.}{{Brinkman}
  et~al.}{2000}]{brinkman_2000}
{Brinkman}, A.~C., et~al. 2000, \apjl, 530, L111

\bibitem[\protect\citeauthoryear{{Bromage} et~al.}{{Bromage}
  et~al.}{1992}]{bromage_1992}
{Bromage}, G.~E., {Kellett}, B.~J., {Jeffries}, R.~D., {Innis}, J.~L.,
  {Matthews}, L., {Anders}, G.~J.,  \& {Coates}, D.~W. 1992, in Astronomical
  Society of the Pacific Conference Series, Vol.~26, Cool Stars, Stellar
  Systems, and the Sun, ed. M.~S. {Giampapa} \& J.~A. {Bookbinder}, 80

\bibitem[\protect\citeauthoryear{{Brown} et~al.}{{Brown}
  et~al.}{1998}]{brown_1998}
{Brown}, G.~V., {Beiersdorfer}, P., {Liedahl}, D.~A., {Widmann}, K.,  \&
  {Kahn}, S.~M. 1998, \apj, 502, 1015

\bibitem[\protect\citeauthoryear{{Chambliss}}{{Chambliss}}{1976}]{chambliss_1976}
{Chambliss}, C.~R. 1976, \pasp, 88, 762

\bibitem[\protect\citeauthoryear{{Cutispoto}}{{Cutispoto}}{1990}]{cutispoto_1990}
{Cutispoto}, G. 1990, \aaps, 84, 397

\bibitem[\protect\citeauthoryear{{Del Zanna}, {Landini}, \& {Mason}}{{Del
  Zanna} et~al.}{2002}]{del_2002}
{Del Zanna}, G., {Landini}, M.,  \& {Mason}, H.~E. 2002, \aap, 385, 968

\bibitem[\protect\citeauthoryear{{Drake} \& {Kashyap}}{{Drake} \&
  {Kashyap}}{2001}]{drake_2001}
{Drake}, J.~J.,  \& {Kashyap}, V. 2001, \apj, 547, 428

\bibitem[\protect\citeauthoryear{{Drake} et~al.}{{Drake}
  et~al.}{2014}]{drake_2014}
{Drake}, J.~J., {Ratzlaff}, P., {Kashyap}, V., {Huenemoerder}, D.~P.,
  {Wargelin}, B.~J.,  \& {Pease}, D.~O. 2014, \apj, 783, 2

\bibitem[\protect\citeauthoryear{{Drake} et~al.}{{Drake}
  et~al.}{2015}]{drake_2015}
{Drake}, S.~A., {Osten}, R.~A., {Krimm}, H., {De Pasquale}, M., {Gehrels}, N.,
  \& {Barthelmy}, S. 2015, The Astronomer's Telegram, 6940, 1

\bibitem[\protect\citeauthoryear{{Dunstone} et~al.}{{Dunstone}
  et~al.}{2008}]{dunstone_2008_2}
{Dunstone}, N.~J., {Hussain}, G.~A.~J., {Collier Cameron}, A., {Marsden},
  S.~C., {Jardine}, M., {Barnes}, J.~R., {Ramirez Velez}, J.~C.,  \& {Donati},
  J.-F. 2008, \mnras, 387, 1525

\bibitem[\protect\citeauthoryear{{Dupree} et~al.}{{Dupree}
  et~al.}{2012}]{dupree_2012}
{Dupree}, A.~K., et~al. 2012, \apj, 750, 73

\bibitem[\protect\citeauthoryear{{Flaccomio} et~al.}{{Flaccomio}
  et~al.}{2010}]{flaccomio_2010}
{Flaccomio}, E., {Micela}, G., {Favata}, F.,  \& {Alencar}, S.~P.~H. 2010,
  \aap, 516, L8

\bibitem[\protect\citeauthoryear{{Fuhrmeister} et~al.}{{Fuhrmeister}
  et~al.}{2011}]{lalitha_2011}
{Fuhrmeister}, B., {Lalitha}, S., {Poppenhaeger}, K., {Rudolf}, N., {Liefke},
  C., {Reiners}, A., {Schmitt}, J.~H.~M.~M.,  \& {Ness}, J.-U. 2011, \aap, 534,
  A133

\bibitem[\protect\citeauthoryear{{Gabriel} \& {Jordan}}{{Gabriel} \&
  {Jordan}}{1969}]{Gabriel_Jordan}
{Gabriel}, A.~H.,  \& {Jordan}, C. 1969, \mnras, 145, 241

\bibitem[\protect\citeauthoryear{{Garc{\'{\i}}a-Alvarez}
  et~al.}{{Garc{\'{\i}}a-Alvarez} et~al.}{2008}]{garcia_2008}
{Garc{\'{\i}}a-Alvarez}, D., {Drake}, J.~J., {Kashyap}, V.~L., {Lin}, L.,  \&
  {Ball}, B. 2008, \apj, 679, 1509

\bibitem[\protect\citeauthoryear{{Garmire} et~al.}{{Garmire}
  et~al.}{2003}]{garmire_2003}
{Garmire}, G.~P., {Bautz}, M.~W., {Ford}, P.~G., {Nousek}, J.~A.,  \& {Ricker},
  G.~R., Jr. 2003, in Society of Photo-Optical Instrumentation Engineers (SPIE)
  Conference Series, Vol. 4851, X-Ray and Gamma-Ray Telescopes and Instruments
  for Astronomy., ed. J.~E. {Truemper} \& H.~D. {Tananbaum}, 28

\bibitem[\protect\citeauthoryear{{Getman} et~al.}{{Getman}
  et~al.}{2008}]{getman_2008}
{Getman}, K.~V., {Feigelson}, E.~D., {Broos}, P.~S., {Micela}, G.,  \&
  {Garmire}, G.~P. 2008, \apj, 688, 418

\bibitem[\protect\citeauthoryear{{Grevesse} \& {Sauval}}{{Grevesse} \&
  {Sauval}}{1998}]{grevesse_1998}
{Grevesse}, N.,  \& {Sauval}, A.~J. 1998, \ssr, 85, 161

\bibitem[\protect\citeauthoryear{{G{\"u}del} et~al.}{{G{\"u}del}
  et~al.}{2001}]{guedel_2001}
{G{\"u}del}, M., et~al. 2001, \aap, 365, L336

\bibitem[\protect\citeauthoryear{{G{\"u}del} et~al.}{{G{\"u}del}
  et~al.}{2007}]{guedel_2007}
{G{\"u}del}, M., et~al. 2007, \aap, 468, 353

\bibitem[\protect\citeauthoryear{{G{\"u}del}, {Guinan}, \&
  {Skinner}}{{G{\"u}del} et~al.}{1997}]{guedel_1997}
{G{\"u}del}, M., {Guinan}, E.~F.,  \& {Skinner}, S.~L. 1997, \apj, 483, 947

\bibitem[\protect\citeauthoryear{{G{\"u}del} \& {Naz{\'e}}}{{G{\"u}del} \&
  {Naz{\'e}}}{2009}]{guedel_2009}
{G{\"u}del}, M.,  \& {Naz{\'e}}, Y. 2009, \aapr, 17, 309

\bibitem[\protect\citeauthoryear{{G{\"u}del} et~al.}{{G{\"u}del}
  et~al.}{2005}]{gudel_2005}
{G{\"u}del}, M., {Skinner}, S.~L., {Briggs}, K.~R., {Audard}, M., {Arzner}, K.,
   \& {Telleschi}, A. 2005, \apjl, 626, L53

\bibitem[\protect\citeauthoryear{{Guirado} et~al.}{{Guirado}
  et~al.}{2011}]{guirado_2011}
{Guirado}, J.~C., {Marcaide}, J.~M., {Mart{\'{\i}}-Vidal}, I., {Le Bouquin},
  J.-B., {Close}, L.~M., {Cotton}, W.~D.,  \& {Montalb{\'a}n}, J. 2011, \aap,
  533, A106

\bibitem[\protect\citeauthoryear{{Hall}}{{Hall}}{1980}]{hall_1976}
{Hall}, D.~S. 1980, \apjl, 236, L137

\bibitem[\protect\citeauthoryear{{Huenemoerder}}{{Huenemoerder}}{2002}]{huenermode_2002}
{Huenemoerder}, D.~P. 2002, in High Resolution X-ray Spectroscopy with
  XMM-Newton and Chandra, ed. G.~{Branduardi-Raymont}, 18

\bibitem[\protect\citeauthoryear{{Huenemoerder} et~al.}{{Huenemoerder}
  et~al.}{2004}]{huenemoerder_2004}
{Huenemoerder}, D.~P., {Boroson}, B., {Schulz}, N.~S., {Canizares}, C.~R.,
  {Buzasi}, D.~L., {Preston}, H.~L.,  \& {Kastner}, J.~H. 2004, in IAU
  Symposium, Vol. 219, Stars as Suns : Activity, Evolution and Planets, ed.
  A.~K. {Dupree} \& A.~O. {Benz}, 238

\bibitem[\protect\citeauthoryear{{Huenemoerder}, {Canizares}, \&
  {Schulz}}{{Huenemoerder} et~al.}{2001}]{huenermoeder_2001}
{Huenemoerder}, D.~P., {Canizares}, C.~R.,  \& {Schulz}, N.~S. 2001, \apj, 559,
  1135

\bibitem[\protect\citeauthoryear{{Kashyap} \& {Drake}}{{Kashyap} \&
  {Drake}}{1998}]{kashyap_1998}
{Kashyap}, V.,  \& {Drake}, J.~J. 1998, \apj, 503, 450

\bibitem[\protect\citeauthoryear{{Kastner} et~al.}{{Kastner}
  et~al.}{2004}]{kastner_2004}
{Kastner}, J.~H., {Huenemoerder}, D.~P., {Schulz}, N.~S., {Canizares}, C.~R.,
  {Li}, J.,  \& {Weintraub}, D.~A. 2004, \apjl, 605, L49

\bibitem[\protect\citeauthoryear{{Kastner} et~al.}{{Kastner}
  et~al.}{2002}]{kastner_2002}
{Kastner}, J.~H., {Huenemoerder}, D.~P., {Schulz}, N.~S., {Canizares}, C.~R.,
  \& {Weintraub}, D.~A. 2002, \apj, 567, 434

\bibitem[\protect\citeauthoryear{{Kastner} et~al.}{{Kastner}
  et~al.}{1997}]{kastner_1997}
{Kastner}, J.~H., {Zuckerman}, B., {Weintraub}, D.~A.,  \& {Forveille}, T.
  1997, Science, 277, 67

\bibitem[\protect\citeauthoryear{{Lalitha} et~al.}{{Lalitha}
  et~al.}{2013}]{lalitha_2013}
{Lalitha}, S., {Fuhrmeister}, B., {Wolter}, U., {Schmitt}, J.~H.~M.~M.,
  {Engels}, D.,  \& {Wieringa}, M.~H. 2013, \aap

\bibitem[\protect\citeauthoryear{{Lalitha} \& {Schmitt}}{{Lalitha} \&
  {Schmitt}}{2013}]{lalitha_2013b}
{Lalitha}, S.,  \& {Schmitt}, J.~H.~M.~M. 2013, \aap

\bibitem[\protect\citeauthoryear{{Lanza} et~al.}{{Lanza}
  et~al.}{1998}]{lanza_1998}
{Lanza}, A.~F., {Catalano}, S., {Cutispoto}, G., {Pagano}, I.,  \& {Rodono}, M.
  1998, \aap, 332, 541

\bibitem[\protect\citeauthoryear{{Little-Marenin} et~al.}{{Little-Marenin}
  et~al.}{1986}]{little_1986}
{Little-Marenin}, I.~R., {Simon}, T., {Ayres}, T.~R., {Cohen}, N.~L.,
  {Feldman}, P.~A., {Linsky}, J.~L., {Little}, S.~J.,  \& {Lyons}, R. 1986,
  \apj, 303, 780

\bibitem[\protect\citeauthoryear{{Martin} \& {Brandner}}{{Martin} \&
  {Brandner}}{1995}]{martin_1995}
{Martin}, E.~L.,  \& {Brandner}, W. 1995, \aap, 294, 744

\bibitem[\protect\citeauthoryear{{Mewe} et~al.}{{Mewe}
  et~al.}{1995}]{mewe_1995}
{Mewe}, R., {Kaastra}, J.~S., {Schrijver}, C.~J., {van den Oord}, G.~H.~J.,  \&
  {Alkemade}, F.~J.~M. 1995, \aap, 296, 477

\bibitem[\protect\citeauthoryear{{Ness} \& {Schmitt}}{{Ness} \&
  {Schmitt}}{2005}]{ness_2005}
{Ness}, J.-U.,  \& {Schmitt}, J.~H.~M.~M. 2005, \aap, 444, L41

\bibitem[\protect\citeauthoryear{{Pasquini} et~al.}{{Pasquini}
  et~al.}{1991}]{pasquini_1991}
{Pasquini}, L., {Cutispoto}, G., {Gratton}, R.,  \& {Mayor}, M. 1991, \aap,
  248, 72

\bibitem[\protect\citeauthoryear{{Pasquini} et~al.}{{Pasquini}
  et~al.}{1989}]{pasquini_1989}
{Pasquini}, L., {Schmitt}, J.~H.~M.~M., {Harnden}, F.~R., Jr., {Tozzi}, G.~P.,
  \& {Krautter}, J. 1989, \aap, 218, 187

\bibitem[\protect\citeauthoryear{{Popper} \& {Ulrich}}{{Popper} \&
  {Ulrich}}{1977}]{popper_1977}
{Popper}, D.~M.,  \& {Ulrich}, R.~K. 1977, \apjl, 212, L131

\bibitem[\protect\citeauthoryear{{Porquet} \& {Dubau}}{{Porquet} \&
  {Dubau}}{2000}]{porquet_2000}
{Porquet}, D.,  \& {Dubau}, J. 2000, \aaps, 143, 495

\bibitem[\protect\citeauthoryear{{Pradhan} \& {Shull}}{{Pradhan} \&
  {Shull}}{1981}]{Pradhan}
{Pradhan}, A.~K.,  \& {Shull}, J.~M. 1981, \apj, 249, 821

\bibitem[\protect\citeauthoryear{{Prato} et~al.}{{Prato}
  et~al.}{2001}]{prato_2001}
{Prato}, L., et~al. 2001, \apj, 549, 590

\bibitem[\protect\citeauthoryear{{Protassov} et~al.}{{Protassov}
  et~al.}{2002}]{protassov_2002}
{Protassov}, R., {van Dyk}, D.~A., {Connors}, A., {Kashyap}, V.~L.,  \&
  {Siemiginowska}, A. 2002, \apj, 571, 545

\bibitem[\protect\citeauthoryear{{Randich}}{{Randich}}{1998}]{randich_1998}
{Randich}, S. 1998, in Astronomical Society of the Pacific Conference Series,
  Vol. 154, Cool Stars, Stellar Systems, and the Sun, ed. R.~A. {Donahue} \&
  J.~A. {Bookbinder}, 501

\bibitem[\protect\citeauthoryear{{Reale} et~al.}{{Reale}
  et~al.}{1997}]{reale_1997}
{Reale}, F., {Betta}, R., {Peres}, G., {Serio}, S.,  \& {McTiernan}, J. 1997,
  \aap, 325, 782

\bibitem[\protect\citeauthoryear{{Robrade} \& {Schmitt}}{{Robrade} \&
  {Schmitt}}{2005}]{robrade_2005}
{Robrade}, J.,  \& {Schmitt}, J.~H.~M.~M. 2005, \aap, 435, 1073

\bibitem[\protect\citeauthoryear{{Rodono} et~al.}{{Rodono}
  et~al.}{1986}]{rodono_1986}
{Rodono}, M., et~al. 1986, \aap, 165, 135

\bibitem[\protect\citeauthoryear{{Rucinski}}{{Rucinski}}{1982}]{rucinski_1982}
{Rucinski}, S.~M. 1982, Information Bulletin on Variable Stars, 2203, 1

\bibitem[\protect\citeauthoryear{{Saba} et~al.}{{Saba}
  et~al.}{1999}]{saba_1999}
{Saba}, J.~L.~R., {Schmelz}, J.~T., {Bhatia}, A.~K.,  \& {Strong}, K.~T. 1999,
  \apj, 510, 1064

\bibitem[\protect\citeauthoryear{{Sanz-Forcada}, {Maggio}, \&
  {Micela}}{{Sanz-Forcada} et~al.}{2003}]{sanz_2003}
{Sanz-Forcada}, J., {Maggio}, A.,  \& {Micela}, G. 2003, \aap, 408, 1087

\bibitem[\protect\citeauthoryear{{Schrijver} et~al.}{{Schrijver}
  et~al.}{1995}]{schrijver_1995}
{Schrijver}, C.~J., {Mewe}, R., {van den Oord}, G.~H.~J.,  \& {Kaastra}, J.~S.
  1995, \aap, 302, 438

\bibitem[\protect\citeauthoryear{{Setiawan} et~al.}{{Setiawan}
  et~al.}{2008}]{setiawan_2008}
{Setiawan}, J., {Henning}, T., {Launhardt}, R., {M{\"u}ller}, A., {Weise}, P.,
  \& {K{\"u}rster}, M. 2008, \nat, 451, 38

\bibitem[\protect\citeauthoryear{{Singh}, {Drake}, \& {White}}{{Singh}
  et~al.}{1996}]{singh_1996b}
{Singh}, K.~P., {Drake}, S.~A.,  \& {White}, N.~E. 1996, \aj, 111, 2415

\bibitem[\protect\citeauthoryear{{Singh}, {White}, \& {Drake}}{{Singh}
  et~al.}{1996}]{singh_1996a}
{Singh}, K.~P., {White}, N.~E.,  \& {Drake}, S.~A. 1996, \apj, 456, 766

\bibitem[\protect\citeauthoryear{{Siviero}, {Dallaporta}, \&
  {Munari}}{{Siviero} et~al.}{2006}]{siviero_2006}
{Siviero}, A., {Dallaporta}, S.,  \& {Munari}, U. 2006, Baltic Astronomy, 15,
  387

\bibitem[\protect\citeauthoryear{{Skinner}, {Audard}, \& {G{\"u}del}}{{Skinner}
  et~al.}{2011}]{skinner_2011}
{Skinner}, S.~L., {Audard}, M.,  \& {G{\"u}del}, M. 2011, \apj, 737, 19

\bibitem[\protect\citeauthoryear{{Smith} et~al.}{{Smith}
  et~al.}{2001}]{smith_2001}
{Smith}, R.~K., {Brickhouse}, N.~S., {Liedahl}, D.~A.,  \& {Raymond}, J.~C.
  2001, \apjl, 556, L91

\bibitem[\protect\citeauthoryear{{Soderblom} et~al.}{{Soderblom}
  et~al.}{1998}]{soderblom_1998}
{Soderblom}, D.~R., et~al. 1998, \apj, 498, 385

\bibitem[\protect\citeauthoryear{{Stelzer} et~al.}{{Stelzer}
  et~al.}{2005}]{stelzer_2005}
{Stelzer}, B., {Flaccomio}, E., {Montmerle}, T., {Micela}, G., {Sciortino}, S.,
  {Favata}, F., {Preibisch}, T.,  \& {Feigelson}, E.~D. 2005, \apjs, 160, 557

\bibitem[\protect\citeauthoryear{{Stelzer} et~al.}{{Stelzer}
  et~al.}{2009}]{stelzer_2009}
{Stelzer}, B., {Hubrig}, S., {Orlando}, S., {Micela}, G., {Mikul{\'a}{\v s}ek},
  Z.,  \& {Sch{\"o}ller}, M. 2009, \aap, 499, 529

\bibitem[\protect\citeauthoryear{{Strassmeier} \& {Rice}}{{Strassmeier} \&
  {Rice}}{2000}]{strassmeier_2000}
{Strassmeier}, K.~G.,  \& {Rice}, J.~B. 2000, \aap, 360, 1019

\bibitem[\protect\citeauthoryear{{Telleschi} et~al.}{{Telleschi}
  et~al.}{2005}]{telleschi_2005}
{Telleschi}, A., {G{\"u}del}, M., {Briggs}, K., {Audard}, M., {Ness}, J.-U.,
  \& {Skinner}, S.~L. 2005, \apj, 622, 653

\bibitem[\protect\citeauthoryear{{Tokovinin}}{{Tokovinin}}{1999}]{tokovinin_1999}
{Tokovinin}, A.~A. 1999, Astronomy Letters, 25, 669

\bibitem[\protect\citeauthoryear{{Walter} \& {Bowyer}}{{Walter} \&
  {Bowyer}}{1981}]{walter_1981}
{Walter}, F.~M.,  \& {Bowyer}, S. 1981, \apj, 245, 671

\bibitem[\protect\citeauthoryear{{Walter} et~al.}{{Walter}
  et~al.}{1980}]{walter_1980}
{Walter}, F.~M., {Bowyer}, S., {Linsky}, J.~L.,  \& {Garmire}, G. 1980, \apjl,
  236, L137

\bibitem[\protect\citeauthoryear{{Webb} et~al.}{{Webb}
  et~al.}{1999}]{webb_1999}
{Webb}, R.~A., {Zuckerman}, B., {Platais}, I., {Patience}, J., {White}, R.~J.,
  {Schwartz}, M.~J.,  \& {McCarthy}, C. 1999, \apjl, 512, L63

\bibitem[\protect\citeauthoryear{{Weisskopf} et~al.}{{Weisskopf}
  et~al.}{2002}]{weisskopf_2002}
{Weisskopf}, M.~C., {Brinkman}, B., {Canizares}, C., {Garmire}, G., {Murray},
  S.,  \& {Van Speybroeck}, L.~P. 2002, \pasp, 114, 1

\end{thebibliography}

\appendix

\begin{table*}[!ht]
\tiny
\centering
\caption{Line identification and line fluxes for the entire HD 155555~AB spectra and HD~155555~C spectra.}
\begin{tabular}{cccccccccccc}
\tableline

&&&&~~~~~~~~~~~~~~~~~~~~~~HD~155555~AB&&~~~~~~~~~~~~HD~155555~C&\\
\tableline 
\colhead{Ion} & \colhead{log T$_{max}$} & \colhead{Transition } & \colhead{ $\lambda_{pred}$} & \colhead{$\lambda_{obs}$} &
\colhead{F$_{l}$} & \colhead{$\lambda_{obs}$}& \colhead{F$_{l}$} \\
      & [K] 		& [upper $\rightarrow$ lower]& [$\AA$]	& [$\AA$]		&  [counts] & [$\AA$]& [counts] & \\
\tableline
\tableline

\ion{Ar}{18}     		   & 7.60       & Unknown													& 3.727 &  3.727 & 35.93$\pm$7.05       	& ...     & ...             \\
\ion{Ar}{16}  			   & 7.20  		& Unknown													& 3.952 &  3.951 & 52.68$\pm$8.30   		& ...     & ...             \\
\ion{Ar}{16}/XVII  & 7.20/7.30  & Unknown												 	& 3.994 &  3.994 & 46.76$\pm$7.89   		& ...     & ...             \\
\ion{S}{16}   & 7.40  & $2p~~^2P_{3/2}\rightarrow 1s~~^2S_{1/2}$    								& 4.727 &  4.731 & 77.71$\pm$9.85 		    & 4.732	  & 3.34$\pm$3.02	\\ 
\ion{S}{15}   & 7.20  & $1s~~2p~~^1 P_{1}\rightarrow 1s^2~~^1S_{0}$									& 5.038 &  5.041 & 147.70$\pm$13.18     	& 5.036   & 9.18$\pm$4.15	\\
\ion{Si}{14}  & 7.20  & $2p~~^2P_{3/2}\rightarrow 1s~~^2S_{1/2}$ 									& 6.180 &  6.184 & 409.75$\pm$21.26 	    & 6.185   & 20.85$\pm$5.64  \\ 
\ion{Si}{13}  & 7.00  & $1s~~2p~~^2P_{2}\rightarrow 1s^2~~^1S_{0}$									& 6.650 &  6.649 & 579.86$\pm$25.09 	    & 6.652   & 28.23$\pm$6.38  \\ 
\ion{Si}{13}  & 6.95  & $1s~~2p~~^3P_{2}\rightarrow 1s^2~~^1S_{0}$									& 6.688 &  6.688 & 52.49$\pm$8.29		 	& ...     & ...			    \\
\ion{Si}{13}  & 7.00  & $1s~~2s~~^3S_{1}\rightarrow 1s^2~~^1S_{0}$ 									& 6.740 &  6.740 & 371.91$\pm$20.30 	    & 6.740   & 9.63$\pm$4.22  	\\
\ion{Mg}{12}  & 7.00  &	$3p~~^2P_{3/2}\rightarrow 1s~~^2S_{1/2}$									& 7.110 &  7.109 & 191.56$\pm$14.86  		& ...     & ...             \\
\ion{Fe}{24}  & 7.30  &	$1s^2~~5p~~^2P_{3/2}\rightarrow 1s^2~~^2s~~^2S_{1/2}$						& 7.170 &  7.168 & 193.70$\pm$14.94 		& ...     & ...             \\
\ion{Mg}{11}  & 6.85  & Unknown																		& 7.473 &  7.471 & 132.60$\pm$12.54       	& ...     & ...             \\
\ion{Al}{12}  & 6.90  &	$1s~~2p~~^1P_{1}\rightarrow 1s^2~~^1S_{0}$									& 7.760 &  7.757 & 166.77$\pm$13.94  		& 7.757   & 6.88$\pm$3.71       \\
\ion{Mg}{11}  & 6.80  & $1s~~3p~~^1P_{1}\rightarrow 1s^2~~^1S_{0}$									& 7.850 &  7.852 & 186.06$\pm$14.66  		& ...  	  & ...				\\ 
\ion{Fe}{24}  & 7.29  & $1s^2~~4p~~^2P_{3/2}\rightarrow 1s^2 2s~~^2S_{1/2}$ 						& 7.985 &  7.986 & 211.71$\pm$15.57  		& 7.980	  & 13.53$\pm$4.77				\\ 
\ion{Fe}{23}  & 7.20  &	$2s~~4p~~^1P_{1}\rightarrow 2s^2~~^1S_{0}$									& 8.300 &  8.310 & 261.50$\pm$17.19  		& ...     & ...            \\
\ion{Mg}{12}  & 7.00  & $2p~~^2P_{3/2}\rightarrow 1s~~^1S_{1/2}$ 									& 8.419 &  8.420 & 1030.06$\pm$33.10 		& 8.422   & 36.62$\pm$7.11		\\ 
\ion{Fe}{23}  & 7.20  & $2s~~4d~~^1D_{2}\rightarrow 2s~~2p~~^1P_{1}$								& 8.810 &  8.812 & 196.60$\pm$15.04  		& ...     & ...         \\
\ion{Fe}{22}  & 7.10  &	$2s^2~~4d~~^2D_{3/2}\rightarrow 2s^2~~2p~~^2P_{1/2}$						& 8.980 &  8.973 & 149.39$\pm$13.25  		& ...     & ...           \\
\ion{Mg}{11}  & 6.81  & $1s~~2p~~^1P_{1}\rightarrow 1s^2~~^1S_{0}$									& 9.168 &  9.169 & 651.05$\pm$23.45 	    & 9.168   & 15.59$\pm$5.04   	\\
 \ion{Mg}{11} & 6.81  & $1s~~2p~~^3P_{2}\rightarrow 1s^2~~^1S_{0}$									& 9.231	&  9.234 & 162.23$\pm$13.76 	    & ...	  & ...				\\ 
\ion{Mg}{11}  & 6.81  & $1s~~2s~~^3S_{2}\rightarrow 1s^2~~^1S_{0}$									& 9.314 &  9.314 & 316.62$\pm$18.81 	    & 9.310	  &26.30$\pm$6.20				\\ 
\ion{Fe}{21}  & 7.10  &	$2s^2~~2p~~4d~~^3D_{1}\rightarrow 2s^2~~2p^2~~^3P_{0}$						& 9.480 &  9.479 & 112.90$\pm$11.66     	& ...     &   ...          \\
\ion{Fe}{21}  & 7.10       & Unknown													            & 9.542 &  9.545 & 37.14$\pm$7.15       	& ...     & ...             \\
\ion{Ne}{10}  & 6.81  &	$4p~~^2P_{3/2}\rightarrow 1s~~^2S_{1/2}$ 									& 9.708 &  9.707 & 269.70$\pm$17.44 		& ...	  &	...				\\ 
\ion{Ni}{25}  & 7.30  &	$2s~~3d~~^1D_{2}\rightarrow 2s~~2p~~^1P_{1}$			    				& 9.970 &  9.993 & 170.32$\pm$14.07  		& ...      &   ...        \\
\ion{Na}{11}  & 6.90  &	$2p~~^2P_{1/2}\rightarrow 1s~~^2S_{1/2}$				    				& 10.020& 10.038 & 264.61$\pm$17.29  		& ...      &   ...         \\
\ion{Ne}{10}  & 6.78  &	$3p~~^2P_{3/2}\rightarrow 1s~~^2S_{1/2}$ 					                &10.238 & 10.239 & 566.43$\pm$24.81 		&10.242 &	15.22$\pm$4.99	\\ 
\ion{Fe}{25}  & 7.75       & Unknown													            & 10.369 & 10.362 & 142.44$\pm$12.96       	& ...     & ...             \\
\ion{Fe}{24}  & 7.27  & $1s^2~~3p~~^2P_{3/2}\rightarrow 1s^2~~2s~~^2S_{1/2}$		                &10.619 & 10.625 & 287.08$\pm$17.96 		&10.616 & 11.42$\pm$4.48		\\ 
\ion{Fe}{24}  & 7.27  & $1s^2~~3p~~^2P_{1/2}\rightarrow 1s^2~~2s~~^2S_{1/2}$ 		                &10.663 & 10.664 & 186.58$\pm$14.68     	&10.663	  & 16.04$\pm$5.09				\\ 
\ion{Ne}{9}   & 6.60  & Unknown																		&10.764 & 10.768 & 120.82$\pm$12.02       	& ...     & ...             \\
\ion{Fe}{23}? & 7.2   & Unknown													                    & 10.825& 10.817 & 114.83$\pm$11.75       	& ...     & ...             \\
\ion{Fe}{23}  & 7.20  &	$2s~~3p~~^1P_{1}\rightarrow 2s^2~~^1S_{0}$					                & 10.980& 10.984 & 211.56$\pm$15.57  		&10.989    & 9.81$\pm$4.24         \\
\ion{Fe}{23}  & 7.20  & $2p~~3p~~^3P_{1}\rightarrow 2s^2~~^1S_{0}$           						&11.020 & 11.028 & 265.94$\pm$17.33			&11.028 &15.75$\pm$5.06		\\
\ion{Fe}{24}  & 7.29  & $1s^2~~3d~~^2D_{5/2}\rightarrow 1s^2~~2p~~^2P_{3/2}$ 						&11.176 & 11.177 & 249.64$\pm$16.82 			&11.176 &11.76$\pm$4.53		\\ 
\ion{Fe}{24}  & 7.29  & Unknown																& 11.260 & 11.259 & 139.02$\pm$12.69       	& ...     & ...             \\
\ion{Fe}{18}  & 6.90  & Unknown														&11.326 & 11.321 & 98.98$\pm$10.98					&... 	  &...				\\ 
\ion{Fe}{24}  & 7.29  & $1s^2~~3s~~^2S_{1/2}\rightarrow 1s^2~~2p~~^2P_{3/2}$ 		&11.432 & 11.430 & 302.00$\pm$18.39 			&11.430	 &10.97$\pm$4.42				\\ 
\ion{Ne}{9}   & 6.60  & $1s~~3p~~^1P_{1}\rightarrow 1s^2~~^1S_{0}$ 					&11.544 & 11.539 & 250.20$\pm$16.84 			&... 	  &...				\\ 
\ion{Fe}{23}  & 7.18  & $2s~~3d~~^1D_{2}\rightarrow 2s~~2p~~^1P_{1}$ 				&11.736 & 11.741 & 298.14$\pm$18.28 			&...	  &...				\\ 
\ion{Fe}{23}  & 7.10  & $2S^2~~3d~~^2D_{3/2}\rightarrow 2s^2~~2p~~^2P_{1/2}$ 		&11.770	& 11.775 & 308.51$\pm$18.58     		&11.753   &21.03$\pm$5.66	\\ 
\ion{Fe}{22}  & 7.10  &	$2s^2~~3d~~^2D_{5/2}\rightarrow 2s^2~~2p~~^2P_{3/2}$		&11.930 & 11.929 & 111.54$\pm$11.59  		&11.932   &14.38$\pm$4.88         \\
\ion{Fe}{21}/XXII  & 7.1/7.1   &	Unknown														&11.975 & 11.976 & 72.97$\pm$9.58  		&   &  \\
\ion{Ne}{10}  & 6.77  & $2p~~^2P_{3/2}\rightarrow 1s~~2S_{1/2}$						&12.132 & 12.134 & 2099.93$\pm$46.83 		&12.134   &53.21$\pm$8.34	\\ 
\ion{Fe}{21}  & 7.10  & $2s^2~~2p~~3d~~^3D_{1}\rightarrow 2s^2~~2p^2~~^3P_{0}$ 		&12.284 & 12.284 & 470.20$\pm$22.70 			&12.282   &20.38$\pm$5.59	\\ 
\ion{Fe}{21}  & 7.10       & Unknown														& 12.397 &  12.398 & 108.70$\pm$11.46       	& ...     & ...             \\
\ion{Fe}{21}   & 7.10       & Unknown														& 12.649 &  12.650 & 135.44$\pm$12.67       	& ...     & ...             \\
\ion{Fe}{20}  & 7.00       & Unknown														& 12.750&  12.754 & 134.06$\pm$12.61       	& ...     & ...             \\
\ion{Fe}{20}  & 7.00  & $2s^2~~2p^2~~3d~~^4P_{5/2}\rightarrow 2s^2~~2p^3~~^4S_{3/2}$&12.827 & 12.827 & 451.24$\pm$22.26 			&12.840   &17.36$\pm$5.25	\\ 
\ion{Fe}{20}  & 7.00  & Unknown													& 13.143 & 13.143 & 109.68$\pm$11.50       	& ...     & ...             \\
\ion{Ne}{9}   & 6.59  & $1s~~2p~~^1P_{1}\rightarrow 1s^2~~^1S_{0}$					&13.447 & 13.449 & 503.49$\pm$26.53		&13.453	  &13.63$\pm$4.79				\\ 
\ion{Fe}{21}  & 6.91  & $2s~~2p^2~~3s~~^3P_{0}\rightarrow 2s~~2p^3~~^3D_{1}$		&13.518 & 13.515 & 354.99$\pm$19.86 			&13.551	  &9.25$\pm$4.16				\\ 
\ion{Ne}{9}   & 6.59  & $2p^3~~3d~~^3D_{2}\rightarrow 2s^2~~2p^4~~^3P_{1}$			&13.553 & 13.554 & 109.67$\pm$11.50 	&13.701   &22.28$\pm$5.79				 \\ 
\ion{Ne}{9}   & 6.59  & $1s~~2s~~^3S_{1}\rightarrow 1s^2~~^1S_{0}$					&13.699 & 13.699 & 313.47$\pm$18.72 	&...      &...		 \\ 
\ion{Fe}{18}  & 6.84  & $2p^4~~3d~~^2D_{5/2}\rightarrow 2s^2~~2p^5~~^2P_{3/2}$		&14.208 & 14.206 & 404.66$\pm$21.13 			&14.204   &17.64$\pm$5.28	\\ 
\ion{Fe}{18}  & 6.84  & $2p^4~~3d~~^2S_{1/2}\rightarrow 2s^2~~2p^5~~^2P_{3/2}$		&14.256 & 14.266 & 151.33$\pm$13.33			&14.263	  &11.88$\pm$4.55				\\ 
\ion{Fe}{18}  & 6.89  & $2p^4~~3d~~^2D_{3/2}\rightarrow 2s^2~~2p^5~~^2P_{1/2}$		&14.373 & 14.372 & 238.64$\pm$16.47			&14.382	  &16.81$\pm$5.19				\\ 
\ion{Fe}{18}  & 6.90  &	$2p^4~~3d~~^2F_{5/2}\rightarrow 2s^2~~2p^5~~^2P_{3/2}$		&14.530 & 14.540 & 159.65$\pm$13.66  			& ...      & ...        \\
\ion{Fe}{18}  & 6.90  &	$2p^4~~3d~~^4P_{1/2}\rightarrow 2s^2~~2p^5~~^2P_{3/2}$		&14.580 & 14.585 & 48.67$\pm$8.03  			& ...      & ...         \\
\ion{Fe}{19}  & 6.90  &	$2p^3~~3s~~^3D_{3}\rightarrow 2s^2~~2p^4~~^3P_{2}$			&14.670 & 14.669 & 84.79$\pm$10.24  		& ...      & ...         \\
\ion{Fe}{20}     		   & 7.00      & Unknown										& 14.754 &  14.755 & 63.24$\pm$8.99       	& ...     & ...             \\
\ion{Fe}{20}   ?  		   & 7.00       & Unknown										& 14.826 &  14.823 & 60.14$\pm$8.80       	& ...     & ...             \\
\ion{Fe}{17}  & 6.72  & $2p^5~~3d~~^1P_{1}\rightarrow 2p^6~~^1S_{0}$				&15.014 & 15.014 & 653.29$\pm$26.57 			&  15.013 & 45.75$\pm$7.81	\\
 \ion{Fe}{19} & 6.90  &	$2s~~2p^4~~3s~~^3P_{2}\rightarrow 2s~~2p^5~~^3P_{2}$		&15.080 & 15.081 & 121.42$\pm$12.05 		&  15.083 & 6.21$\pm$3.63         \\
\ion{O}{8}    & 6.51  & $4p~~^2P_{3/2}\rightarrow 1s~~^2S_{1/2}$					&15.176 & 15.175 & 135.37$\pm$12.66		&...	  &...				\\ 
\ion{Fe}{17}  & 6.72  & $2p^5~~3d~~^3D_{1}\rightarrow 2p^6~~^1S_{0}$				&15.261 & 15.263 & 230.56$\pm$16.20 			& 15.266	  &17.65$\pm$5.29				\\  
\ion{Fe}{17}     		   & 6.70       & $2p^5~~3d~~^3P_{1}\rightarrow 2p^6 ~~1S_{0}$										& 15.450 &  15.457 & 79.499$\pm$9.95       	& ...     & ...             \\
\ion{Fe}{20}     		   & 7.10       & $2p^4~~(^1D)~~3s~~^2D_{5/2}\rightarrow 2s^2~~2p^5~~2P_{3/2}$										& 15.626 &  15.626 & 116.25$\pm$11.81       	& ...     & ...             \\
\ion{Fe}{18}     		   & 6.90       & $2p^4~~(^3P)~~3s~~^4P_{3/2}\rightarrow 2s^2~~2p^5~~^2P_{3/2}$				 					& 15.828 &  15.823 & 76.51$\pm$9.79       	& ...     & ...             \\
\ion{Fe}{18}     		   & 6.90       & $2p^4~(^1P)~~3s~~^2D_{3/2}\rightarrow 2s^2~~2p^5~~^2P_{1/2}$										& 15.870 &  15.869 & 78.39$\pm$9.89       	& ...     & ...             \\
\ion{O}{8}    & 6.50  & $3p~~^2P_{3/2}\rightarrow 1s~~2S_{1/2}$	    				&16.005 & 16.007 & 387.09$\pm$20.69			& 16.009  &17.56$\pm$5.27	\\ 
\ion{Fe}{18}  & 6.84  & $2p^4~~3s~~^4P_{5/2}\rightarrow 2s^2~~2p^5~~^2P_{3/2}$		&16.071 & 16.073 & 229.54$\pm$16.17 			&16.076	  &7.37$\pm$3.85				\\ 
\ion{Fe}{19}  & 6.90  &	$2p^3~~3p~~^3P_{2}\rightarrow 2s~~2p^5~~^3P_{2}$			&16.120 & 16.108 & 72.81$\pm$9.57  			&...       & ...         \\
\ion{Fe}{18}  & 6.70  &	$2s~~2p^5~~3s~~^2P_{3/2}\rightarrow 2s~~2p^6~~^2S_{1/2}$    &16.170 & 16.167 & 58.14$\pm$8.67 			&...       & ...         \\
\ion{Fe}{17}  & 6.71  & $2p^5~~3s~~^3P_{1}\rightarrow 2p^6~~^1S_{0}$				&16.780 & 16.776 & 208.22$\pm$18.57 			&16.773    & 12.55$\pm$4.64				 \\ 
\ion{Fe}{17}  & 6.71  & $2p^5~~3s~~^1P_{1}\rightarrow 2p^6~~^1S_{0}$				&17.051 & 17.052 & 367.09$\pm$20.17			&17.092   & 24.80$\pm$6.05	\\ 
\ion{Fe}{17}  & 6.71  & $2p^5~~3s~~^3P_{2}\rightarrow 2p^6~~^1S_{0}$				&17.096 & 17.096 & 306.81$\pm$18.53 			&...	  &...				\\ 
\ion{O}{8}    & 6.48  & $2p~~^2P_{3/2}\rightarrow 1s~~^2S_{1/2}$					&18.967 & 18.970 & 781.22$\pm$28.96 			&18.970   &27.80$\pm$6.34	\\
\ion{O}{7}    & 6.33  & $1s~~2p~~^1P_{1}\rightarrow 1s^2~~^1S_{0}$					&21.600 & 21.601 & 70.80$\pm$9.45		&...	  &...				\\ 
\ion{O}{7}    & 6.33  & $1s~~2p~~^1P_{1}\rightarrow 1s^2~~^1S_{0}$					&21.800 & 21.804 & 27.55$\pm$6.32		&...	  &...				\\ 
\ion{O}{7}    & 6.33  & $1s~~2p~~^1P_{1}\rightarrow 1s^2~~^1S_{0}$					&22.100 & 22.106 & 31.15$\pm$6.64		&22.100	  &4.49$\pm$3.29				\\ 
\ion{N}{7}    & 6.32  & $2p~~^2P_{3/2}\rightarrow 1s~~^2S_{1/2}$					&24.779 & 24.780 & 57.54$\pm$8.63 			&...	  &...			  \\ 
\tableline 
\end{tabular}

\end{table*}

\end{document}